\documentclass{elsart}

\usepackage{natbib}

\usepackage{graphicx}

\usepackage{amssymb}

\begin{document}

\runauthor{De Santis, Grazian, Fontana \& Santini}
\runtitle{ConvPhot}

\begin{frontmatter}

\title{ConvPhot: A Profile-Matching Algorithm for Precision Photometry}

\author{C. De Santis, A. Grazian, A. Fontana, P. Santini}
\address{INAF - Osservatorio Astronomico Roma Via Frascati, 33 --
          00040 Monte Porzio Catone (RM), Italy}
\ead{desantis,grazian,fontana,santini@oa-roma.inaf.it}

\begin{abstract}

We describe in this paper a new, public software for accurate
``PSF-matched'' multiband photometry for images of different
resolution and depth, that we have named {\tt ConvPhot}, of which we
analyse performances and limitations. It is designed to work when a
high resolution image is available to identify and extract the
objects, and colours or variations in luminosity are to be measured in
another image of lower resolution but comparable depth.  To maximise
the usability of this software, we explicitly use the outputs of the
popular SExtractor code, that is used to extract all objects from the
high resolution ``detection'' image.  The technique adopted by the
code is essentially to convolve each object to the PSF of the lower
resolution ``measure'' image, and to obtain the flux of each object by a
global $\chi^2$ minimisation on such measure image. We remark that no
a priori assumption is done on the shape of the objects. In this
paper we provide a full description of the algorithm, a discussion of
the possible systematic effects involved and the results of a set of
simulations and validation tests that we have performed on real as
well as simulated images.  The source code of {\tt ConvPhot}, written
in C language under the GNU Public License, is released worldwide.
\end{abstract}


\end{frontmatter}

\section{Introduction}

The availability of efficient imagers, operating with good imaging
quality over a large range of wavelengths, has opened a new era in
astronomy. Multi-wavelength imaging surveys have been executed or
planned, and are providing major breakthroughs in many fields of modern
astronomy. These surveys often collect images of different quality and
depth, typically resulting from the combined effort of ground and
space--based facilities.  In this context, the difficulties
originating in the analysis of these often inhomogeneous imaging databases
have hampered the proper exploitation of these data sets, especially
in the field of faint, high redshift galaxies.

On the one side, images at varying wavelengths may provide a surprisingly 
different glimpse of the Universe, with objects fading or emerging from the
background.  On the other side, the image resolution is
usually not constant over the wavelengths, due to different instrument
characteristics. A typical case is the combination of high resolution
HST images with lower resolution images obtained by ground based
telescopes or by the Spitzer Space telescope. In the latter case, the blending
among the objects in the lower resolution images often prevents a full
exploitation of the multicolour informations contained in the data.  

The difficulties involved in the analysis of this kind of data has led
to the development of several techniques. The first emphasis was based
on refinements of the usual detection algorithms \citep{szalay99}.
The algorithm here discussed, that is designed to work especially for
faint galaxies, allows instead to accurately measure colours in
relatively crowded fields, making full use of the spatial and
morphological information contained in the highest quality images.
This approach has already been proposed and used in previous works
\citep{fernandez99,papovich,grazian}: here, we discuss a specific
implementation of the software code, named {\tt ConvPhot}, that we
developed and make publicly available to analyse data with
inhomogeneous image quality.  Although we focus in the following on
multi-wavelength observations, i.e. the case where different images of
the same portion of sky are available in different bands, the approach
followed here can be adopted also for variability studies, where images in
the same band but taken at different epochs are used. In this case,
the highest quality image can be used to identify the objects, and the
magnitude variations of all the objects can be measured with no
systematic effects.

The plan of the paper is the following: in Sect. 2, we outline the
technique adopted; in Sect. 3, we discuss more exhaustively the algorithm,
of which we provide more details. We also describe the problematics 
involving the determination of the isophotal area and the optimisations
we adopted. In Sect. 4 we comment on possible issues of {\tt ConvPhot},
taking into account the cases of blended sources in the {\it detection} image.
In Sect. 5 we describe the systematics that may be cause of inappropriate
results from the algorithm.
In Sect. 6, we discuss the validation tests that we performed on simulated
as well as on real images, discussing in particular the usage of this
software and the future prospects for improving {\tt ConvPhot}.
A brief summary and conclusions of the paper are given in Sect. 7.


\section{The basic technique}

The technique that we discuss here has been described and adopted for
the first time by the Stony-Brook group to optimise the analysis of
the $J$, $H$ and $K$ images of the Hubble Deep Field North (HDF-N).
The method is described in Fernandez--Soto et al. (1999, hereafter
FSLY99) and the catalog obtained has been used in several scientific
analysis of the HDF-N, both by the Stony-Brook group
\citep{lanzetta,phillipps00}, as well as by other groups, including our own
\citep{fontana00,poli01}.  The same method has been adopted by
\citep{papovich,dickinson}, to deal with the similar problems existing
in the HDFs data set and, recently, to derive a photometric catalog of
the GOODS fields in the 24 micron band of MIPS (PSF is 5 arcseconds)
using the 3.6 micron band of IRAC (1.6 arcsec of resolution) to
deblend \citep{chary} the MIPS sources\footnote{\tt
http://data.spitzer.caltech.edu/popular/goods/Documents/goods\_dr3.html}.

Although in both these cases the method has been applied to the
combination between HST and ground--based, near--IR data, the
procedure is much more general and can be applied to any combination
of data sets, provided that the following assumptions are satisfied:\\
{\it 1.} A high resolution image (hereafter named {\it``detection
image''}) is available, that is used to detect objects and isolate
their area;  this image should be deep enough to allow all the
sources in the {\it ``measure image''} to be detected. Ideally, such
image should be well sampled and allow a proper resolution for most
sources.\\
{\it 2.}  Colours (or temporal variations) are to be measured in a
lower resolution image (hereafter named {\it``measure image''});\\
{\it 3.}  The PSF is accurately estimated in both images, and a
convolution kernel has been obtained to smooth the {\it``detection
image''} to the PSF of the {\it``measure image''}.

\begin{figure}[t]
\centering
\includegraphics[width=12cm]{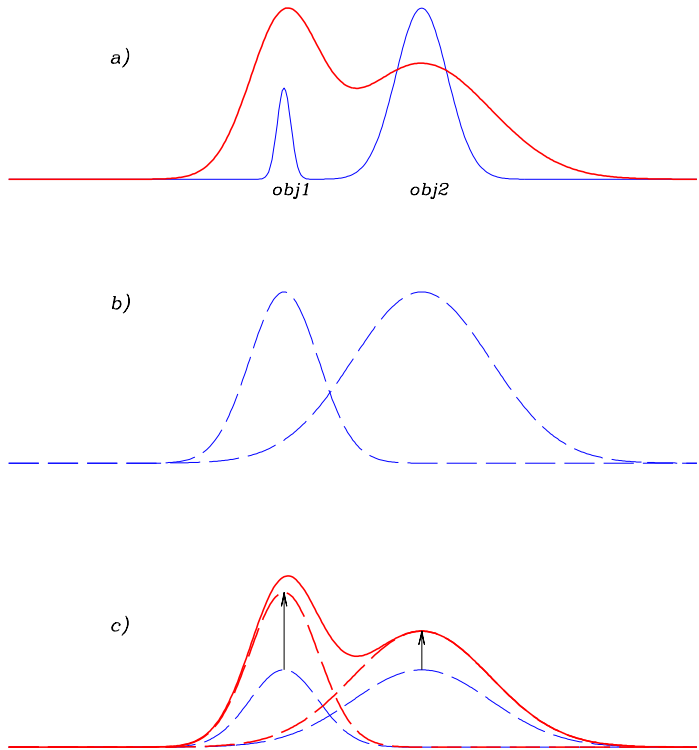}
\caption{A schematic representation of the {\tt ConvPhot} algorithm.
{\it a)}. Two objects are clearly detected and separated in the high 
resolution {\it detection} image (blue, solid-thin line). The same two 
objects are blended in the low resolution {\it measure} image (red, 
solid-thick line) and have quite different colours.
{\it b)}. The two objects are isolated in the high resolution 
{\it detection} image and are individually smoothed to the PSF of the 
{\it measure} image, to obtain the ``model'' images.
{\it c)}. The intensity of each object is scaled to match the global profile
of the {\it measure} image. The scaling factors are found with a global 
$\chi^2$ minimisation over the object areas.}
\label{conv}
\end{figure}

Conceptually, the method is quite straightforward, and can be better
understood by looking at Fig.~\ref{conv}. In the upper panel, we plot the
case of two objects that are clearly detected in the {\it``detection
image''}, but severely blended in the {\it``measure''} one.
The technique described here can remind that adopted by DOPHOT
\citep{dophot}, a software for PSF fitting in crowded stellar fields.
There are otherwise some differences, since {\tt ConvPhot} is thought
especially for galaxies or extended objects in general, and the minimisation
of the model image to the measure image is done simultaneously,
reducing the possibility of wrong fits in severely blended objects, an
usual fact when the fit is carried out into different steps.
Indeed, the {\tt ConvPhot} software needs a model of the convolution kernel
in input, while in DOPHOT the PSF is fitted analytically.

The procedure followed by {\tt ConvPhot} consists of the following steps:\\

{\it a)} Object identification is done relying the parameters and area
obtained by a SExtractor \citep{sextractor} run on the {\it``detection
image''}: in practice, we use the ``Segmentation'' image produced by
SExtractor, that is obtained from the isophotal area. We extract small
images centered around each object, on which all the subsequent
operations will be performed. Such small images (dynamically stored in
the computer's memory) are named ``thumbnails'' in the
following. Details are described in Section~3.1, including the
treatment of border effects.\\

{\it b)} Since the isophotal area is typically an underestimate of the
actual object size, such that a fraction of the object flux is lost in
the tails outside the contour of the last isophote, we have developed a
specific software (named {\tt dilate}) to expand the SExtractor area
of an amount $f_{DIL}$ proportional to the object size. This procedure
is described in Section~3.2 \\

{\it c)} Another required input is a convolution kernel that converts
the detection image to the same resolution of the {\it ``measure''}
image. Although this is not provided within {\tt ConvPhot}, a short
description of the requirements and possible methods is given in
Section 3.3.

{\it d)} The background of each object (both in the {\it``detection''}
and in the {\it ``measure''} images) is computed, as described in Section~3.4;

{\it e)} Each object, along with its segmentation image, is
(individually) smoothed, and then normalised to unit total flux: we
refer to it as the ``model profiles'' of the objects, as it represents
the expected shape of the object in the {\it ``measure''} image
(Section 3.5).\\

{\it f)} Finally, the intensity of all ``model'' objects are
simultaneously adjusted in order to match the intensity of the
objects in the {\it``measure''} image. This global minimisation
provides the final normalisation (i.e. intensity) of each object in
the ``model'' image, and hence the fundamental output of {\tt
ConvPhot}. This is described in Section 3.6

In this scaling procedure there are as many free parameters (that we
named $F_i$, the fitted flux of the $i$-th galaxy) as the number of objects
in the detection image, that are
typically thousands.  The free parameters $F_i$ are computed with a
$\chi^2$ minimisation over all the pixels of the images, and all
objects are fitted simultaneously to take into account the effects of
blending between nearby objects. Conceptually, this approach is
somewhat similar to the line fitting of complex absorption metal
systems in high redshift quasars \citep{fb95}.  Although in practical
cases the number of free parameters can be quite large, the resulting
linear system is very sparse (FSLY99), since non null terms represent
only the rare overlapping/blended sources, and the minimisation can be
performed in a quite efficient way by using standard numerical
techniques.

As can be appreciated from the example plotted in Fig.~\ref{conv}, the
main advantage of the method is that it relies on the accurate spatial
and morphological information contained in the ``detection'' image to
measure colours in relatively crowded fields, even in the case that
the colours of blended objects are markedly different.

Still, the method relies on a few assumptions that must be well
understood and taken into account.  First of all, it is assumed
that the objects have no measurable proper motion. This is not a
particular concern in small, deep extragalactic fields like HDF or
GOODS, but might be important in other applications. A more general
concern is that morphology and positions of the objects should not
change significantly between the two bandwidths. Also, the depth and
central bandpass of the {\it ``detection''} image must ensure that
most of the objects detected in the {\it``measure''} image are
contained in the input catalog. Finally, the objects should be well
separated on the detection image, although they may be blended in the
measure image.

In practice, it is unlikely that all these conditions are satisfied in
real cases. In the case of the match between ACS and ground based $Ks$
images, for instance, very red objects may be detected in the $Ks$
band with no counterpart in the optical images, and some morphological
change is expected due to the increasing contribution of the bulge in
the near--IR bands. This mis-match may be particularly significant for
high redshift sources, where the knots of star formation in the UV
rest frame are redshifted to the optical wavelengths used as model
images.  Also, in the case that the pixel-size of the {\it``measure''}
image (i.e. ISAAC or VIMOS) is much larger than that of the
{\it``input''} one (ACS for example), the actual limitations due to
intrinsic inaccuracy in image aligning may lead to systematic errors.

We will show below that these effects may be minimised or corrected
with reasonable accuracy: at this purpose, we have included in the
code several options and fine--tuning parameters to minimise the
systematics involved.


\section{The ConvPhot algorithm}

In this paragraph, we illustrate in detail the algorithm
adopted for the {\tt ConvPhot} software. Particular emphasis is given
to resolve critical issues in this profile matching algorithm, such
as a correct estimation of the background,
an unbiased reproduction of the object profile and a fast and robust
minimisation technique. We analyse in detail these critical points.

\subsection{Input images and catalog}

As mentioned above, {\tt ConvPhot} relies on the output of the popular
SExtractor code to detect and deblend the objects and to define the
object position and main morphological parameters in the {\em
detection} image.  In the following, we shall also use the SExtractor
naming convention to specify several image characteristics. We remind
that a ``segmentation'' image is an image where background is set to 0
and all the pixels assigned to a given objects are filled with the
catalog number of the object.
 
\begin{figure}[t]
\centering
\includegraphics[width=12cm]{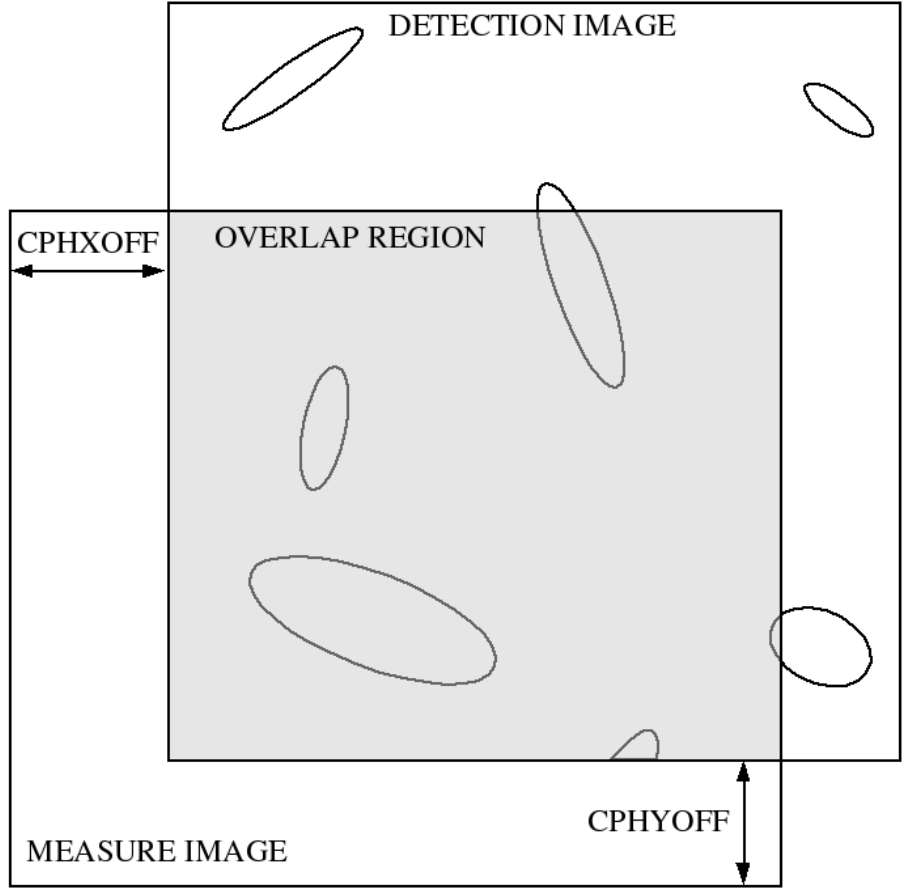}
\caption{This figure shows how {\tt ConvPhot}
defines the overlap region between the 
{\it detection} and the {\it measure} image in case of different sizes and
offsets, 
provided by the input parameters CPHXOFF and CPHYOFF. The
objects 
whose segmentation is partly inside the overlap region are flagged as
{\it bound}.
}
\label{offset}
\end{figure}

The user must first obtain a reliable SExtractor catalog of the {\it
detection} image prior to running {\tt ConvPhot}. In practice, the
required output from SExtractor consists of the SEGMENTATION image  and
of a catalog with the center and boundary coordinates of the objects
in pixel units and the value of the background, as shown in
Tab. \ref{inputcat}.

In addition to this, the user must provide the {\it detection} image
and the {\it measure} with its MAP\_RMS,  i.e. an image containing the
absolute r.m.s. of the {\it measure} image, in the same units. It is
important to remind that the {\it measure} image may be shifted and/or
have a different size with respect to the {\it detection} one: it is
only required that the images are aligned (i.e. have been rebinned to
the same pixel size and be astrometrically registered; for problems related
to the rebinning
of the {\it measure} image see Sect. 6). This offsetting option may be useful
when dealing with the follow--up of large mosaics (i.e. GOODS, COSMOS).
Fig.~\ref{offset} shows the case of a {\it measure} image having a
different size and shifted respect to the {\it detection} one. In this
case, the relative offset (in pixels) between the two images must be
provided by the input parameters CPHXOFF and CPHYOFF. The {\it measure} image 
and its RMS must have the same
size. As shown in Fig.~\ref{offset}, {\tt ConvPhot} works on objects
whose segmentation lies in the overlapping region between the
detection and the measure image. In the {\it detection} image, objects
which partly lie inside the overlap region are flagged as {\it bound}
and will be processed only if the fraction between the flux inside the
overlap region and the total flux in the {\em detection} image is
equal or larger than a user-defined parameter $f_{flux}$.

\begin{table*}
\caption[]{Example of input catalog for {\tt ConvPhot}.}
\label{inputcat}
\begin{tabular}{ccccccccc}
\hline
\hline
NUMBER & X & Y & XMIN & XMAX & YMIN & YMAX & BACK\_D & BACK\_M \\
\hline
1 & 2319.71 & 216.86 & 2134 & 2507 & 34 & 384 & 1.350E-05 & 0.0 \\
2 & 8849.97 & 79.61 & 8820 & 8881 & 32 & 124 & -1.300E-06 & 0.0 \\
3 & 5269.25 & 16.99 & 5245 & 5293 & 1 & 39 & -6.372E-06 & 0.0 \\
4 & 3596.99 & 21.14 & 3581 & 3613 & 5 & 37 & -2.710E-05 & 0.0 \\
5 & 4878.07 & 11.14 & 4852 & 4905 & 1 & 37 & -6.634E-06 & 0.0 \\
\hline
\hline
\end{tabular}
 NUMBER is the identification number of the source in the segmentation
image created by SExtractor, X, Y, XMIN, XMAX, YMIN, YMAX are the center
and boundary coordinates of the objects (in SExtractor they are indicated as
X\_IMAGE, Y\_IMAGE, XMIN\_IMAGE, XMAX\_IMAGE, YMIN\_IMAGE, YMAX\_IMAGE,
respectively). BACK\_D is the value of the background for the detection
image, while BACK\_M is the background for the measure image. It is
possible to provide every object with its local background estimate,
for example that produced by SExtractor. This example is taken from
the GOODS-MUSIC catalog produced using the $z$ band provided by ACS in
the GOODS South field \citep{grazian}.
\end{table*}


\subsection{Small segmentation for objects}

The ``object segmentation'' defines the area (i.e. the set of pixels)
over which each object is distributed. This area, which is different
in the {\it measure} and in the {\it detection} image because of the
typical different image quality (PSF), is used to extract the object
profile in the {\it detection} image and to define the fitting region
in the {\it measure} one.

To establish the area of each object in the {\it detection} image,
{\tt ConvPhot} relies on the ``segmentation'' image produced by
SExtractor. In this image, each pixel contained within the isophotal
area of an object is filled with the relevant object ID, and can be
used to reconstruct the shape of the object itself. This estimate,
however, is not very robust since the size of the isophotal area
depends on the isophotal threshold adopted in the {\it detection}
image, and in any case misses a significant fraction of the flux of
faint objects.

This effect is particularly dramatic as the galaxy becomes either very
faint or very large, as in space based images like those provided by
HST (WFPC2, ACS), with resulting low surface brightness of the
external pixels. To overcome this problem, we have developed and distributed
a dedicated software (named {\tt dilate}), that expand the
segmentation produced by SExtractor, though preventing merging between
neighbouring objects and preserving the original shape of the segmentation.
The segmentation of each object is individually smoothed with an
adaptive kernel which preserves its shape and is then renormalised to the
original number value. This software allows to fix the dilate factor
$f_{DIL}$ for the magnification of the segmentation, that is defined
as the ratio between the output and input isophotal area.  While this
procedure is useful for
relatively bright objects, of which the increase of size is doubled for
an $f_{DIL}=4$,
it provides too small an enlargement for the faintest objects,
that are often detected over very few pixels (this is particularly
true in the case of faint small galaxies observed with HST, as in the
case of the GOODS data). At this purpose, we allow to define a minimum
area $m_{AREA}$ for the segmentation, such that objects smaller than
this area (even after dilation) are forced to have this area.  A
minimum value for the area of extended segmentation is useful, for
instance, in the ACS images, where extended but low surface brightness
objects are detected by SExtractor only due to the bright nucleus and their 
isophotal area is limited only to the brighter knots.  These values
should be tuned accurately, because too small values for $f_{DIL}$ or
$m_{AREA}$ may cause a distortion of the profile for the detection
image, which alters the photometry in the measure image, while too
large values for $f_{DIL}$ or $m_{AREA}$ are not useful since enhance
the noise in the model profile.  Typical values for $f_{DIL}$ are 3-4,
in order to double the area of the original segmentation image. The
$m_{AREA}$ parameter is set in order to match 2-3 times the typical
size of galaxies in the field: for deep imaging surveys with HST,
faint galaxies have an half light radius of $Rh=0.2-0.3$ arcsec, which
translates to a minimum area of around 800 pixels, if the pixel scale
of ACS (0.03 arcsec) is used. Moreover, {\tt ConvPhot} also compute
the so-called ``detection magnitude'', that is the magnitude of the object
in the {\it detection} image for the dilated segmentation. This quantity
should be compared with the total detection magnitude of the
input catalog in order to correct the model magnitude for the fraction of
flux missed by the limited segmentation. In addition, the possibility of
restricting the fit to the central region of the objects (as described in
Sect 3.5) further suppresses the effects of tails in the fit itself.


\subsection{PSF matching}

A key process within {\tt ConvPhot} is the smoothing of the ``{\it
detection}'' high resolution image to the PSF of the lower resolution
``{\it measure}'' one. The original feature of the method is that this
step is not performed on the global image, but rather on each
individual object, after it has been extracted from the ``{\it
detection}'' image, in order to prevent the confusion to to blending
in low resolution images.

Such a smoothing is performed by applying a convolution kernel, that
is the filter required to transform the PSF of the detection image
into the PSF of the ``{\it measure}'' image. Filtering is performed on
the thumbnails of the extracted objects, as well as on the
corresponding segmentation and auxiliary images, whenever required.

The derivation of the convolution kernel is not done by {\tt ConvPhot}
and must be provided by the users.  Several techniques have been
developed to derive an accurate convolution kernel, in particular
those by \cite{alard_lupton} and \cite{alard}, and we advice to use
such sophisticated techniques whenever possible.

In the case of the GOODS--MUSIC sample, where we had to obtain more
than 70 independent kernels (one for each original image of the J,
H and K mosaics), we have adopted a simpler method, based on analysis
in the the Fourier space. Such method can be implemented and
automatized with standard astronomical tools for image analysis.  We
summarise here the basic recipe to compute it. First of all, we derive
the PSF of the detection and measure image, $P_1$ and $P_2$
respectively, e.g. by summing up several stars in the field, in order
to gain in Signal to Noise ratio. These two PSFs must be normalised,
in order to have total flux equal to unity.  The convolution kernel K
is, by definition,
\begin{equation}
P_2(x,y)=K(x,y) \otimes P_1(x,y) \ .
\end{equation}

The derivation of the exact shape of the convolution kernel is done in the
Fourier space: taking ${\aleph} = FT(K)$, ${\wp}_1=TF(P_1)$ and
${\wp}_2=TF(P_2)$, the Fourier transform of the kernel is given by:
\begin{equation}
{\aleph}=\frac{{\wp}_2}{{\wp}_1} \ .
\end{equation}
An optimal Wiener filter (low passband filter),
which suppresses the high frequency fluctuations,
must be applied in the Fourier domain to remove the effects of
noise.  The aim of this filtering is to remove the large frequency
fluctuations due to noise in the two PSFs or with scales less than the pixel
scales of the input images. In order to check for the validity of the used
filter, a residual is computed from the two PSFs and this must be consistent
with the RMS of the coadded PSFs.

Finally, ${\aleph}$ must be anti-transformed back to the pixel space
and the resulting kernel must be normalised to unity, in order to
preserve the flux of the objects.


\subsection{Background determination}

The estimation of an accurate background is essential to determine the
magnitude of faint sources. At this purpose, we have decided to adhere
as much as possible to the SExtractor algorithms to compute the
background of each object, and implemented several options to deal
with this task. In particular, both for the {\it detection} and the
{\it measure} image, the user can select one of these three options:\\

{\it a)} use as input a background--subtracted image, therefore
relying on other applications (such as SExtractor itself) to compute
the background.  This option saves computing time and can be adopted
when the background is quite homogeneous.\\

{\it b)} use the background value computed for each object by
SExtractor, as stored in the input catalog (see
Tab. \ref{inputcat}).\\

\begin{figure}[t]
\centering
\includegraphics[width=3.6cm]{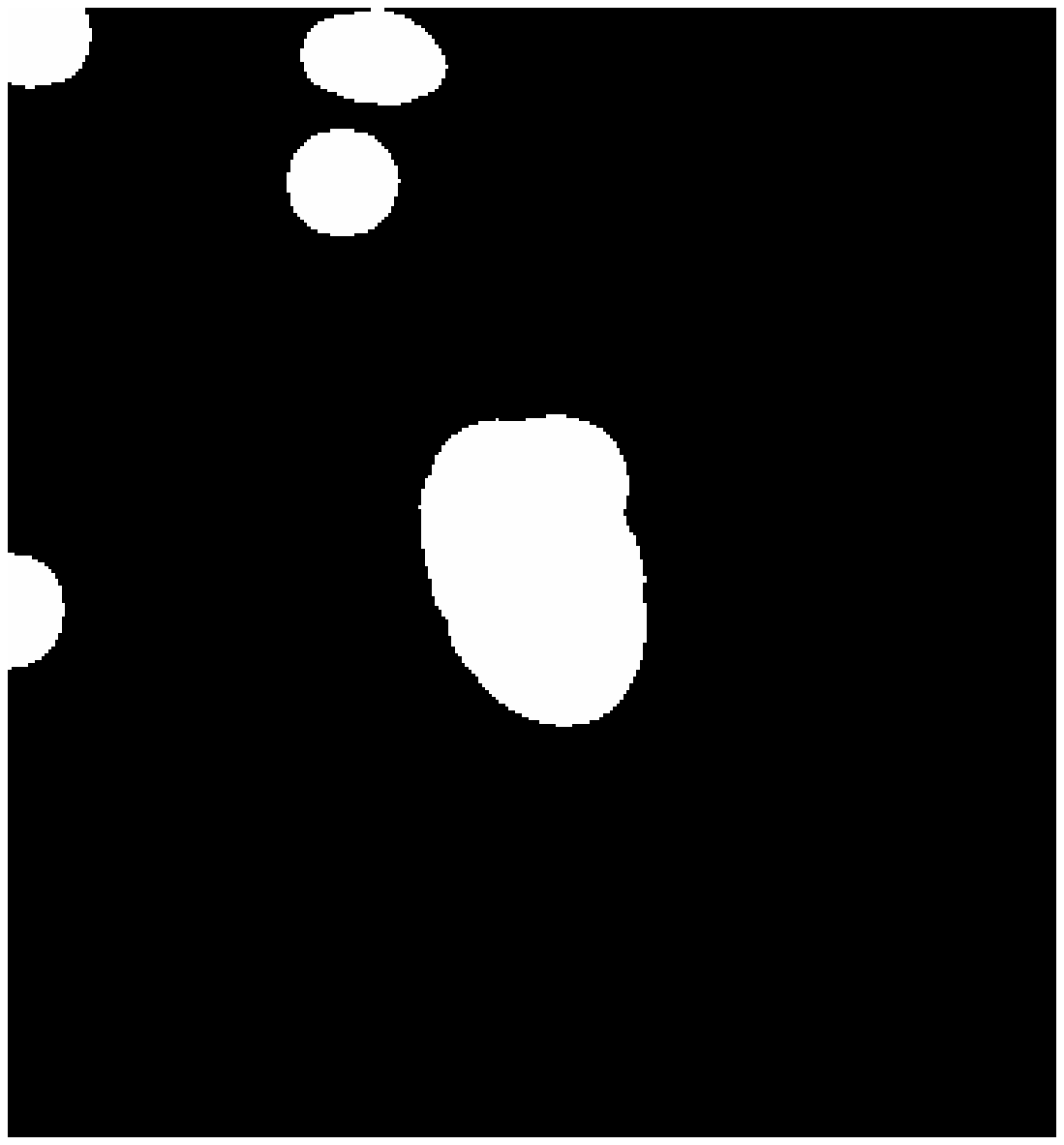}
\includegraphics[width=3.6cm]{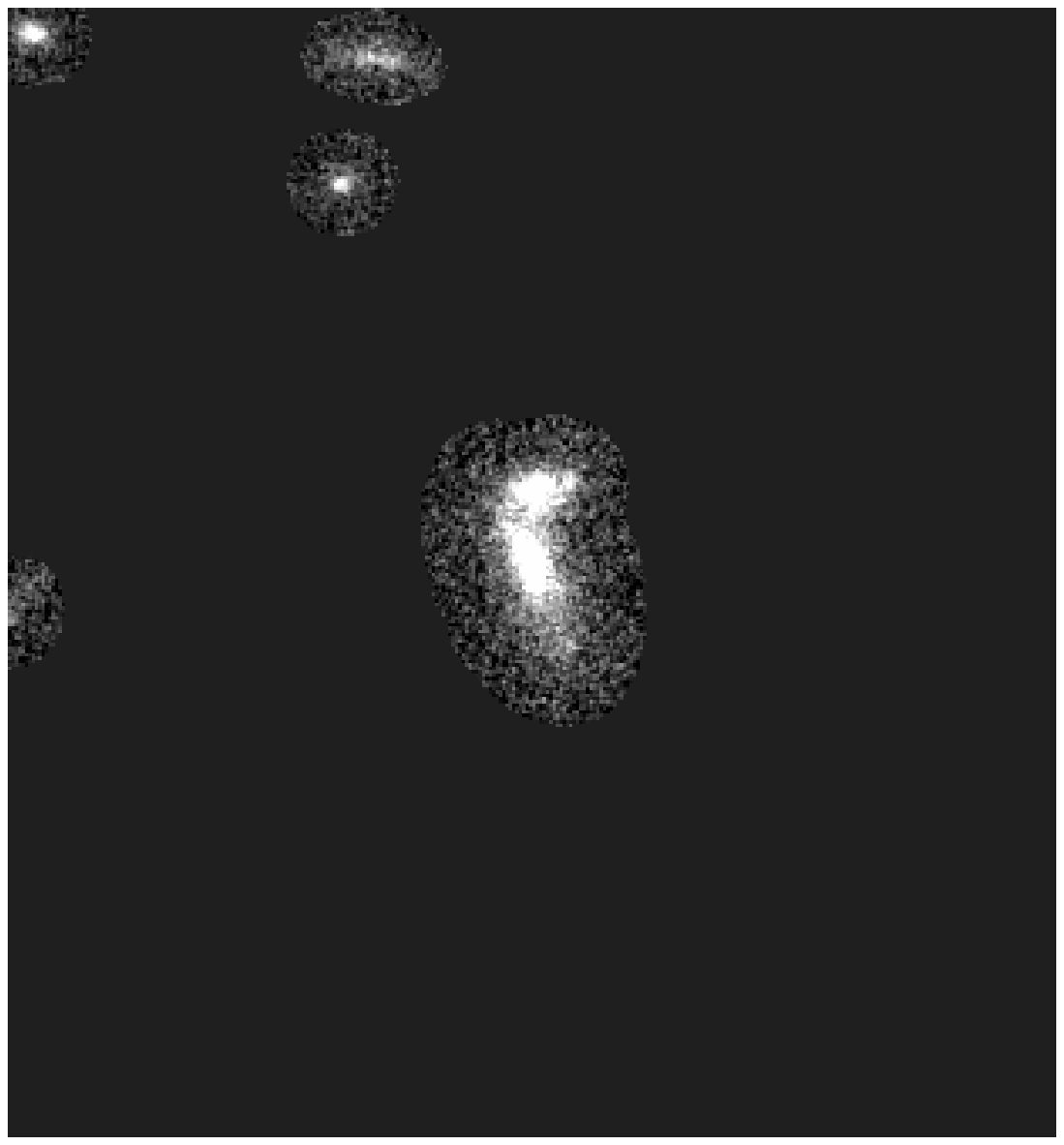}
\includegraphics[width=3.6cm]{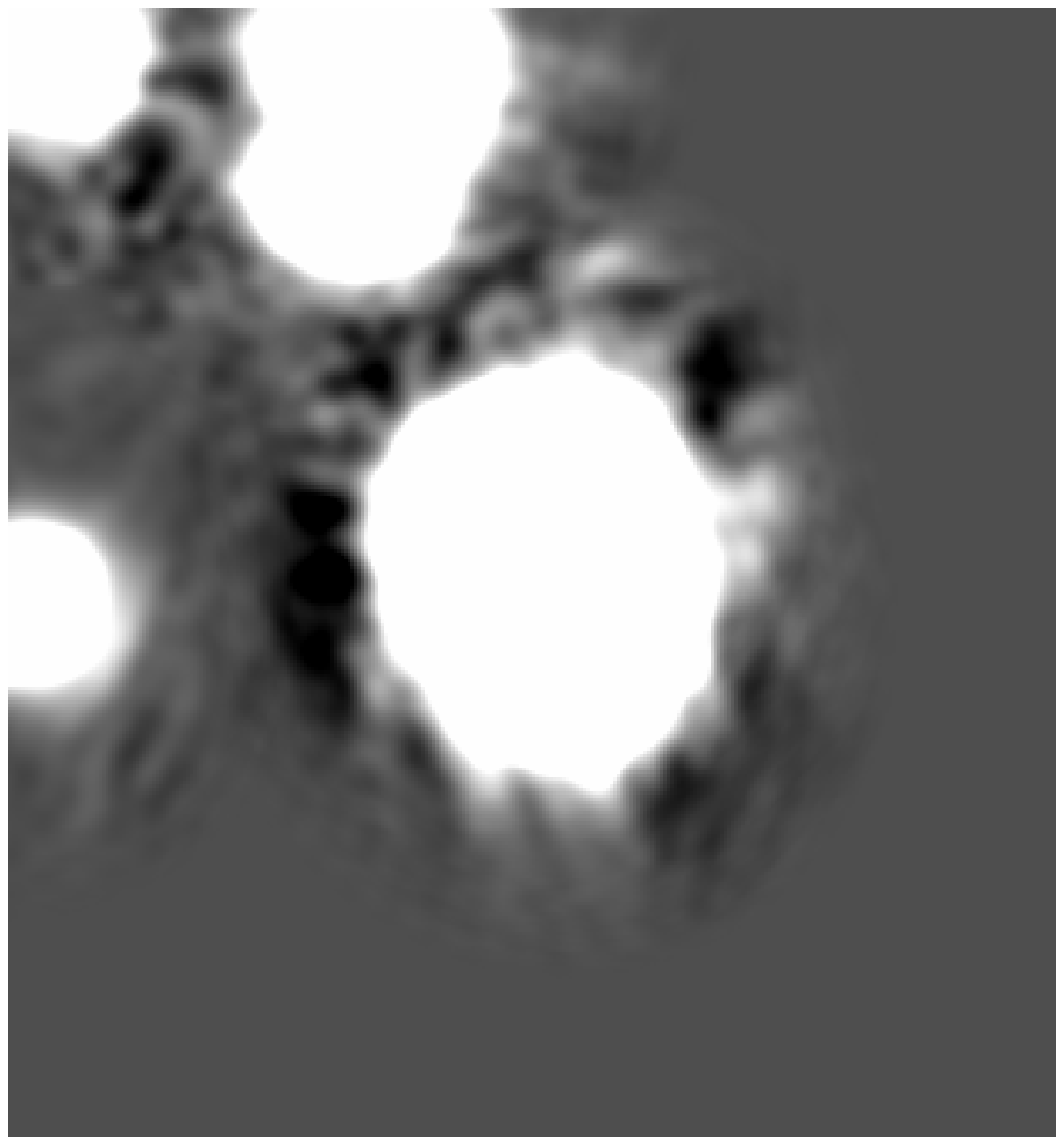}
\includegraphics[width=3.6cm]{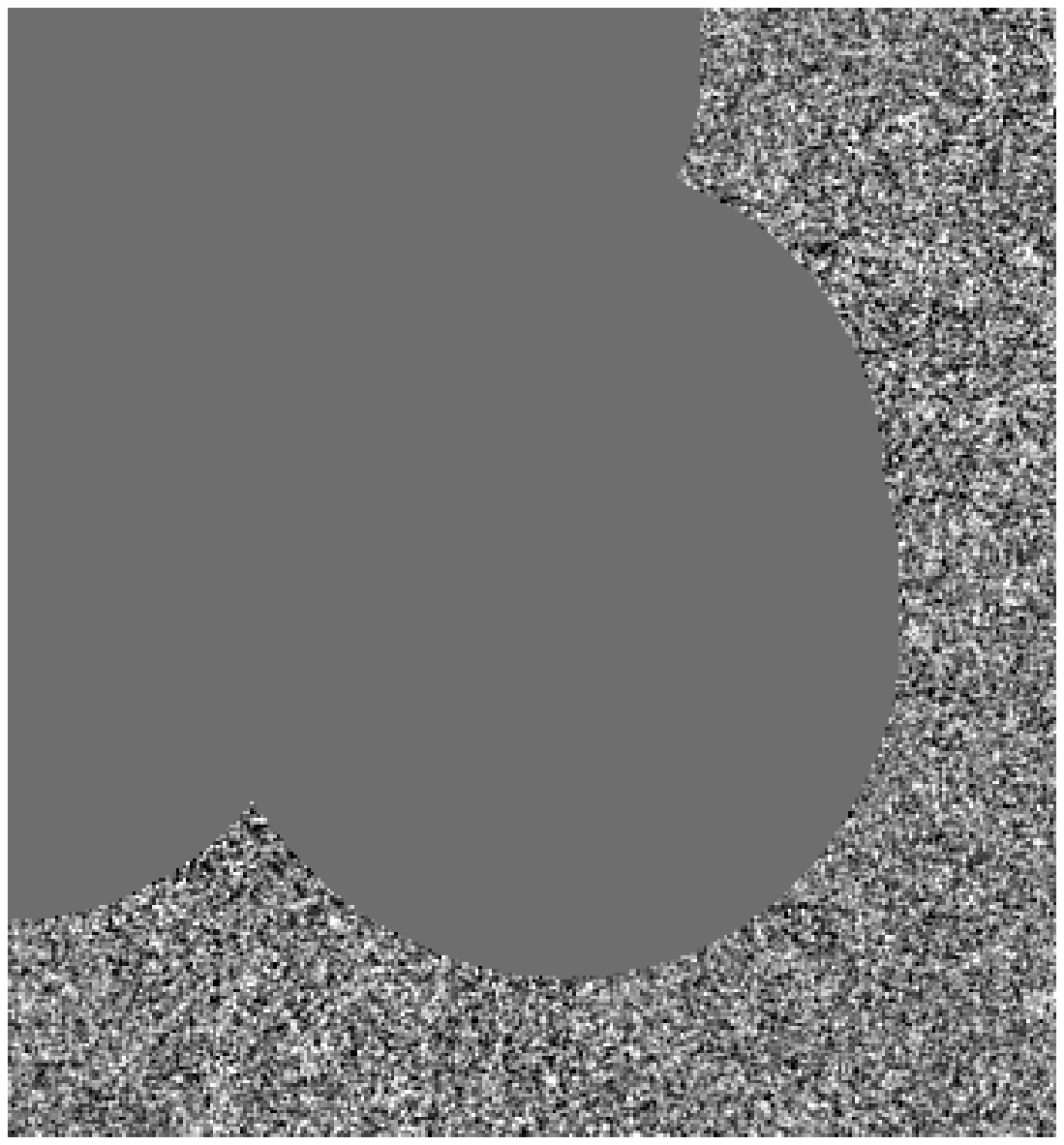}
\includegraphics[width=3.6cm]{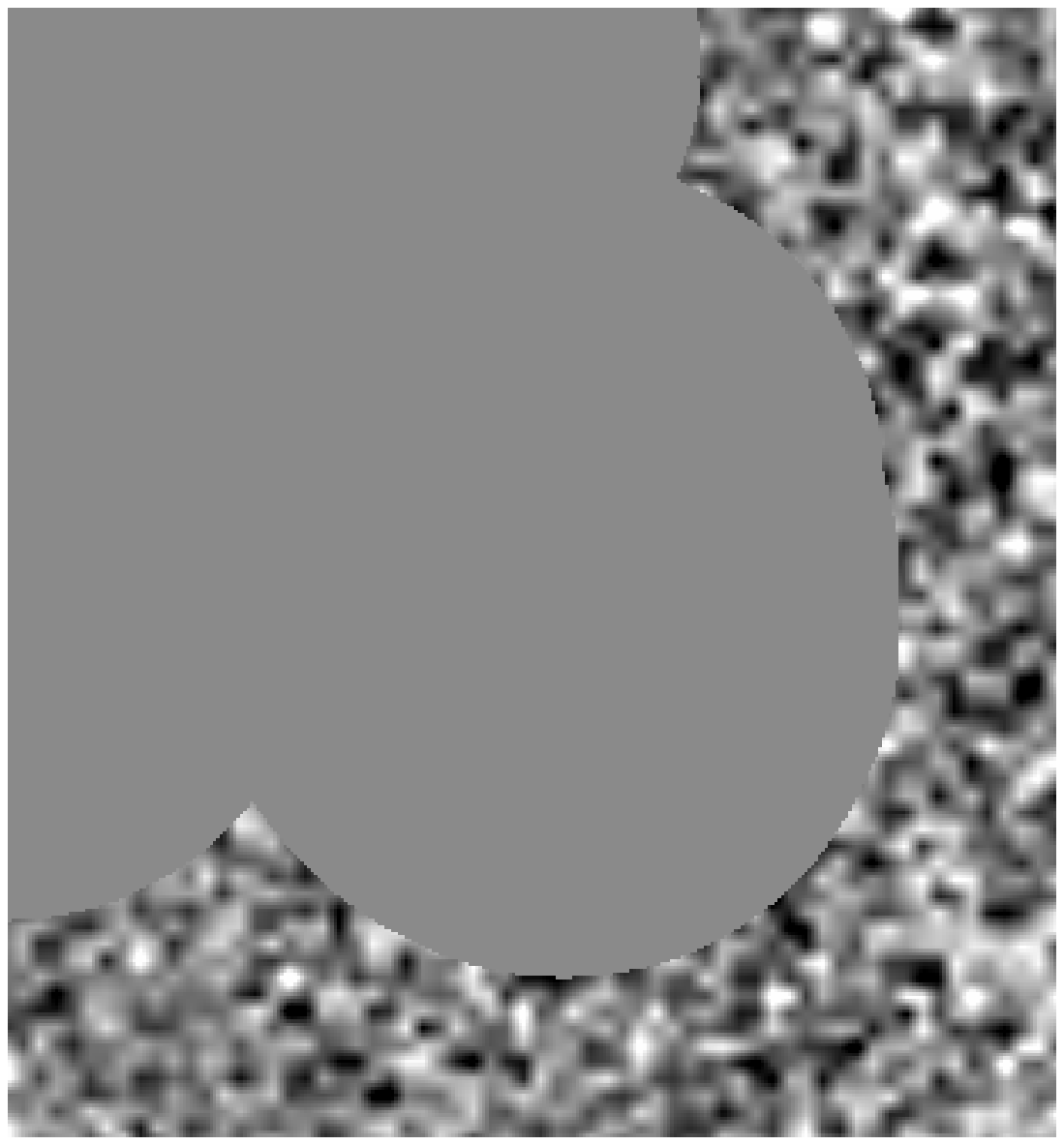}
\includegraphics[width=3.6cm]{fig03F.eps}
\caption{The procedure followed by {\tt ConvPhot}
to obtain the thumbnail of the
local background-area
of a given object. From left to right: {\it a)} the segmentation of the
object and its neighbours 
(white) is extracted, {\it b)} the corresponding objects are extracted
from the {\it detection} image, 
{\it c)} all the objects are smoothed to the {\it measure} PSF, defining the
pixels pertaining to each source, {\it d)}
the ``rectangular annulus'' 
around the object in the {\it detection} image is extracted, excluding those
pixels affected by sources, and here the
BACK\_D value is computed {\it e)} the ``rectangular annulus''
around the object in the {\it measure} image is extracted as in d),
and here the BACK\_M value is computed.
}
\label{background}
\end{figure}

{\it c)} Compute the background within {\tt ConvPhot}: in this case,
the background is computed in a ``rectangular annulus'' around the
object, of size set by the user. Nearby objects are masked out using
the dilated (see Section 3.2) segmentation map (Fig.~\ref{background}).

As a background estimator, in the option c), we follow the SExtractor
approach: first, the program compute the mean and median of the
background pixels with an iterative $\sigma$--clipping method, until
it reaches a convergence at 3 $\sigma$s, then compute the mode as
$MODE=2.5\cdot MEDIAN-1.5\cdot MEAN$ and recompute the RMS of the
background using the mode. If this RMS is changed by more than 20\%
from the previous clipped-RMS, the mean is taken as best estimate for
the background, otherwise {\tt ConvPhot} uses the mode. A detailed
explanation of this approach can be found in Bertin \& Arnouts (1996).
The overall process is described in Fig.\ref{background}.

The $\sigma$--clipping process is essential in rejecting pixels of
possible sources not detected in the detection image but bright in the
measure frame.  At this purpose, the number of pixels used to compute
the background should be larger than a user supplied value, the size
of the ``rectangular annulus''. Thus, the background estimation made
by {\tt ConvPhot} is time consuming since the thumbnails must be much
larger than the typical size of the segmentation for the objects.

\subsection{Model profile}

\begin{figure}[t]
\centering
\includegraphics[width=3.6cm]{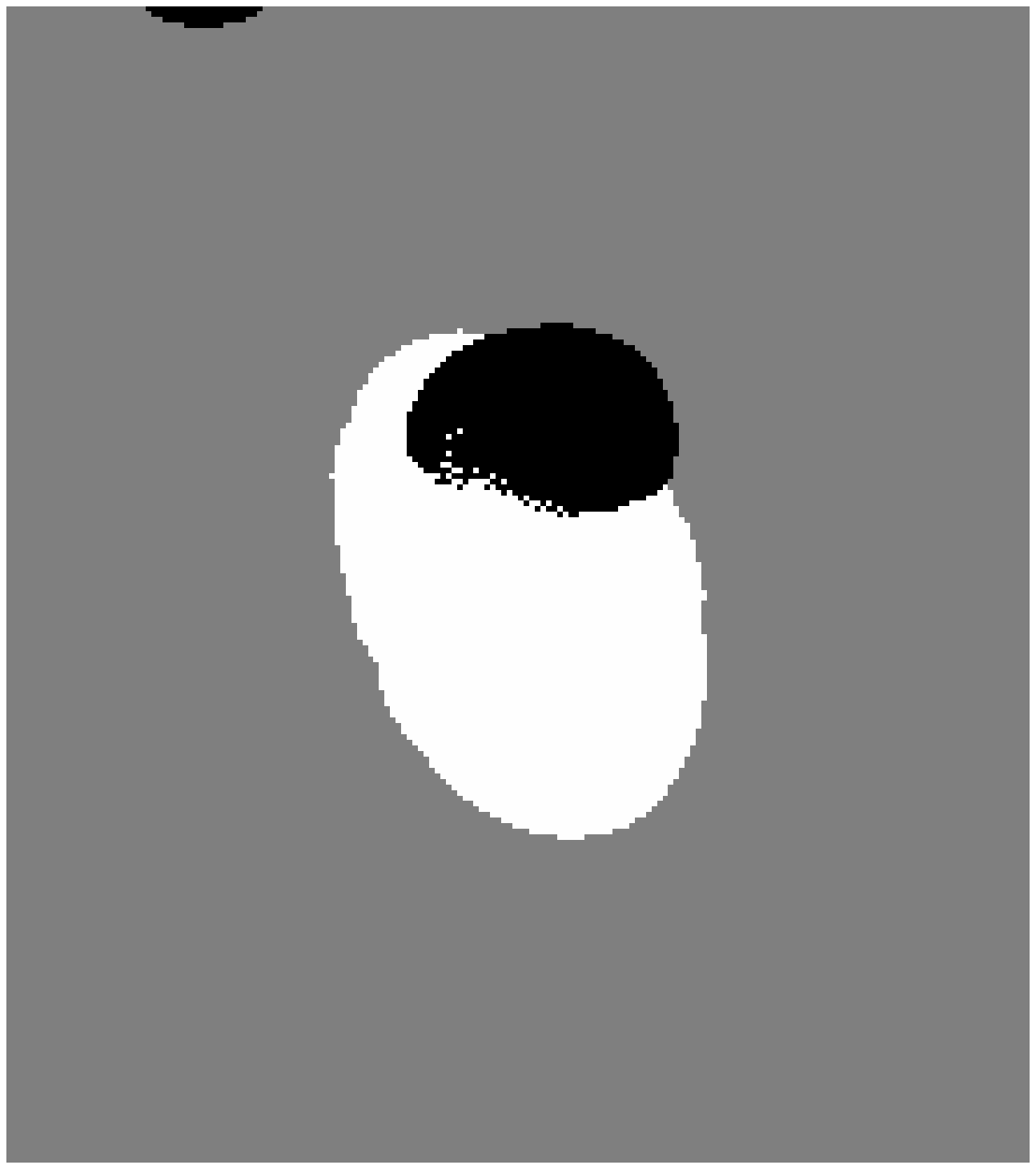}
\includegraphics[width=3.6cm]{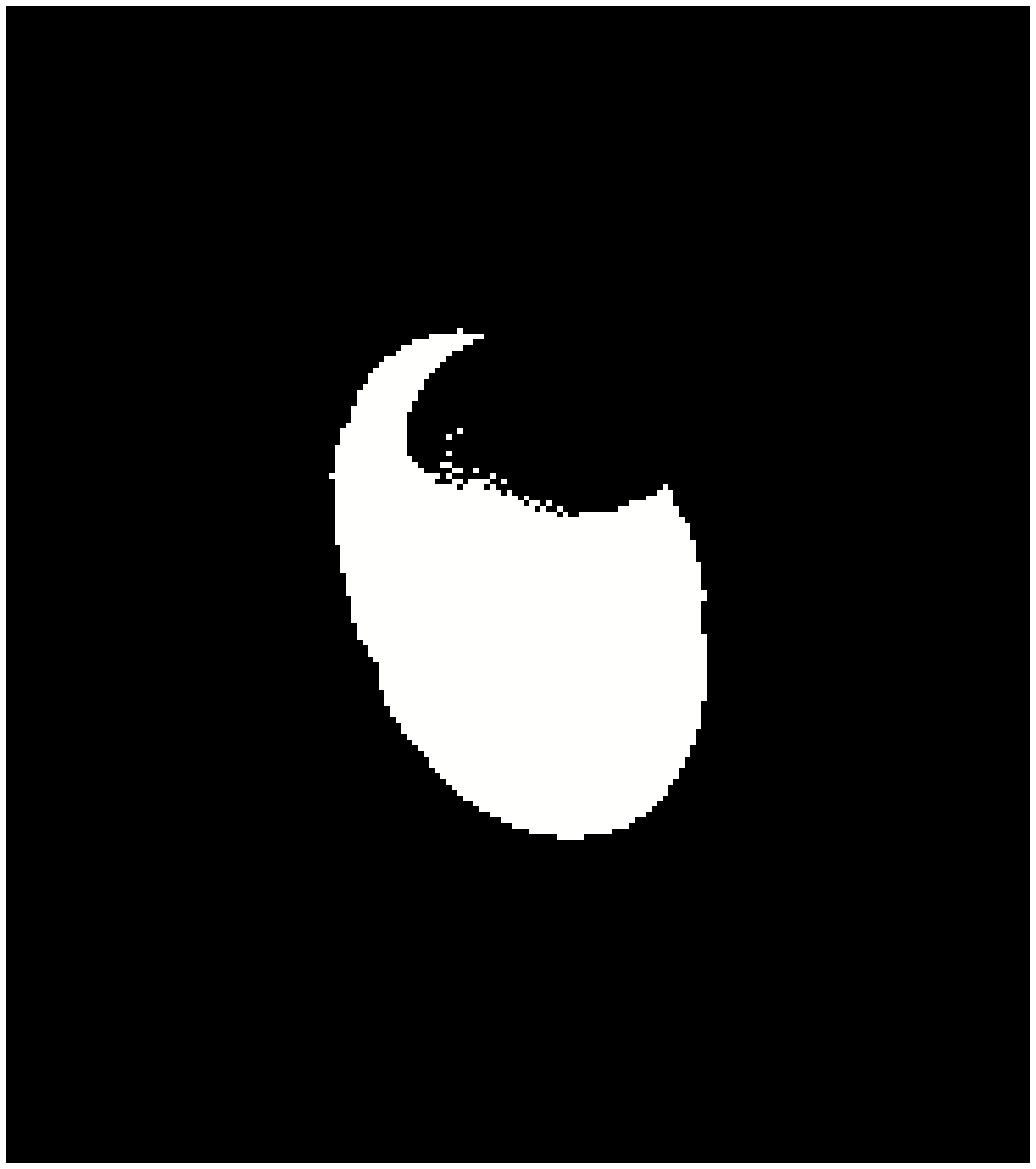}
\includegraphics[width=3.6cm]{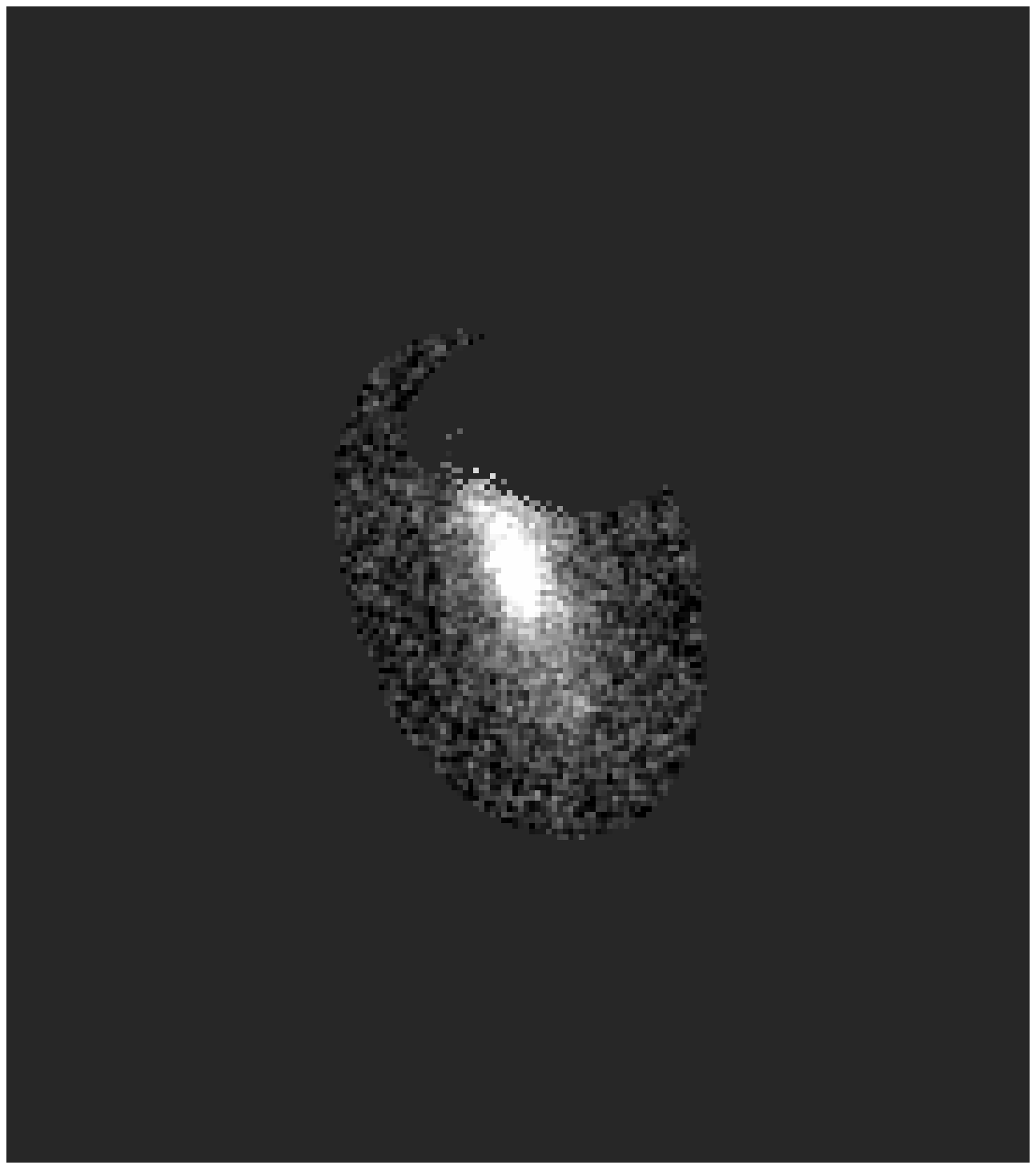}
\includegraphics[width=3.6cm]{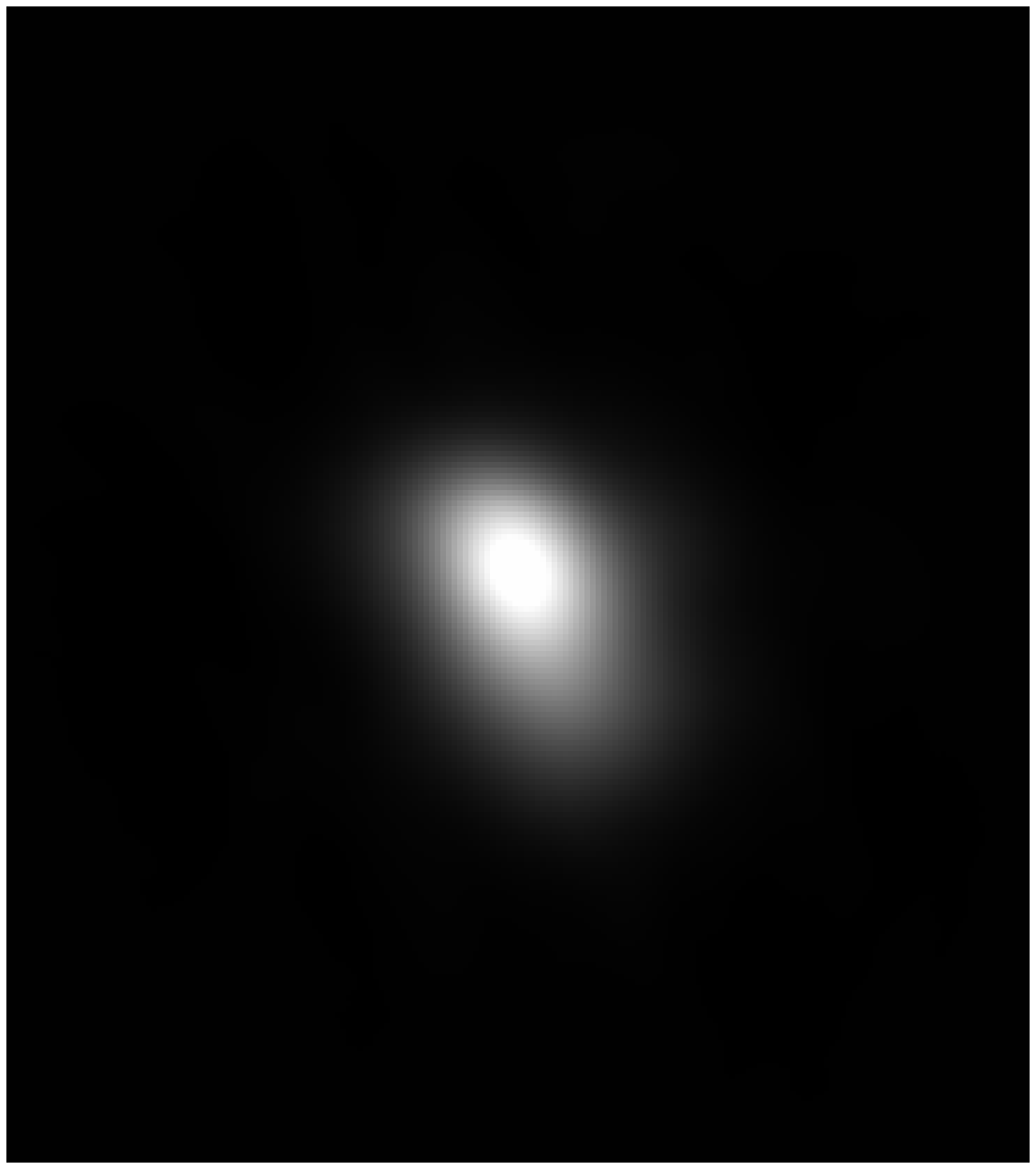}
\includegraphics[width=3.6cm]{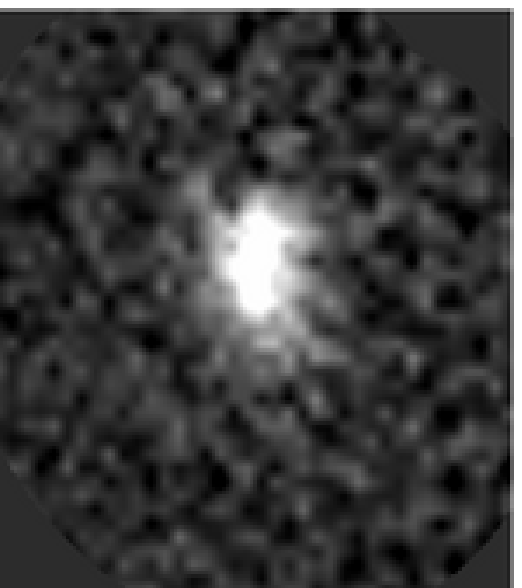}
\includegraphics[width=3.6cm]{fig04F.eps}
\caption{The procedure followed by {\tt ConvPhot} to create the thumbnail of
the {\it model profile} 
of a given object. From left to right: {\it a)} the segmentation of
the object (white) is 
extracted, {\it b)} other objects are masked, {\it c)} the {\it object
profile} is extracted from the 
{\it detection} image and the local background is subtracted, {\it d)}
the {\it object profile} is smoothed 
to the {\it measure} PSF and normalised to obtain the {\it model profile},
{\it e)} the same 
object is extracted from the {\it measure} image subtracting its local
background.
}
\label{thumbnail}
\end{figure}

{\tt ConvPhot} creates a {\it model profile} of each object, whose
segmentation lies inside the overlap region, to measure the colour of
the object itself. In the procedure, depicted in Fig.~\ref{thumbnail},
the (dilated) segmentation of the object is extracted from the {\it
segmentation} image and stored in a thumbnail. The value of the pixels
relative to the given object are set to 1 while the pixels of other
objects and of the background are masked setting them to 0.  Using
this masked segmentation, that defines the isophotal area, the object
is extracted from the {\it detection} image and its background,
provided by the input catalog or estimated by {\tt ConvPhot} itself as
described above, is subtracted producing the {\it object profile}
which is stored in a new thumbnail and smoothed to the seeing of the
{\it measure} image with the PSF--matching kernel. If the object is
flagged as {\it bound} the total flux $D_{i}^{bound}$ is computed and
the thumbnail is resized extracting only the pixels inside the overlap
region (Fig.~\ref{offset}).  Then, the thumbnail isoarea and total
flux $D_{i}$ are computed from the {\it detection} image. For a bound
object, {\tt ConvPhot}
estimates the fraction of total flux $D_{i}/D_{i}^{bound}$ inside the
overlap region and if this value is less than $f_{flux}$ (by default
0.4), it rejects the object.  Finally, the thumbnail is normalised to
unit flux obtaining the {\it model profile} of the object.  The total
flux $D_{i}$ will be used later to obtain self--consistent colours.

For the reasons described below, it may be useful to perform the
fitting procedure only on the central part of the profile: at this
purpose, we have introduced the threshold parameter $t_{f}$, such that
only the pixels above the relative threshold $t_{i}=P_{i}^{max}*t_{f}$
are used, where $P_{i}^{max}$ is the {\it model profile} maximum.
Finally, {\tt ConvPhot} uses the isophotal area of the {\it (selected)
model profile} to extract the same object from the {\it measure} image
subtracting its local background and producing a {\it measure}
thumbnail with the corresponding RMS thumbnail.


\subsection{Minimisation procedure}

We now describe the minimisation process, making explicit use of the
same notation adopted in FSLY99. 

Given the {\it model profiles} (the {\it selected model profiles} 
if the threshold $t_{f}$ has been used) of each object $P_{i}(m,n)$, 
the minimisation procedure aims at obtaining
the scaling factors $F_{i}$ (i.e. the fitted flux of the object in the
measure image assuming as object profile the model image)
of each object that best reproduce
the {\it measure} image $I$.  The best fit solution will be found by 
minimising the $\chi^2$, that reads:
\begin{equation}
\chi^2=  \sum_{m,n}\left[\frac{I(m,n)-B_i-M(m,n)}
{\sigma(m,n)}\right]^2
\end{equation}
where 
\begin{equation}
M (m,n) = \sum_{i=1}^{N_{\rm obj}} F_i P_i(m,n)
\end{equation}
is the sum of all profiles, $B_i$ is the background (in the {\it measure} 
image) fitted for each objects, $\sigma(m,n)$ is the r.m.s. of the
{\it measure} 
image and $m$ and $n$ run over all the pixels.

The best--fit solution is found by solving the linear system:

\begin{equation}
\frac{\partial \chi^2}{\partial F_i}=0 \;\;\;(i=1,\ldots,N_{\rm obj})
\end{equation}

whose Hessian matrix is 

\begin{equation}
A_{ij} = \sum_{m,n}{ \frac{ P_i(m,n) P_j(m,n) } {\sigma(m,n)^2}}
\end{equation}

and right--hand term is:

\begin{equation}
R_i = \sum_{m,n} \frac{ P_i(m,n) [I(m,n)-B_i]} {\sigma(m,n)^2}.
\end{equation}

Since the products of the model profiles are non--zero only within the
area provided
by the (dilated) segmentation map, most of the $A_{ij}$ terms are
actually null (non--null terms will correspond to overlapping
objects), such that the matrix $A$ is very sparse. The solution of
linear systems is performed by an LU decomposition using the
{\it GNU Scientific Library (GSL)} and its {\it LAPACK} routines
(Anderson et al. 1999), such that the
solution can be obtained even for a large number of free parameters
(that is, of detected objects) that typically occur in deep HST
exposures (i.e. UDF, GOODS or COSMOS surveys).

Finally, the statistical uncertainties on the fitted parameters $F_i$
are the diagonal terms of the inverse of the Hessian matrix $A_{ij}$,
which depend on the RMS of the measure image and on the model
profile. In a sense, this fitting procedure is equivalent to derive
a profile-weighted flux and error, relying on the model image.


\subsection{Output quantities}

\begin{figure}[t]
\centering
\includegraphics[width=12cm]{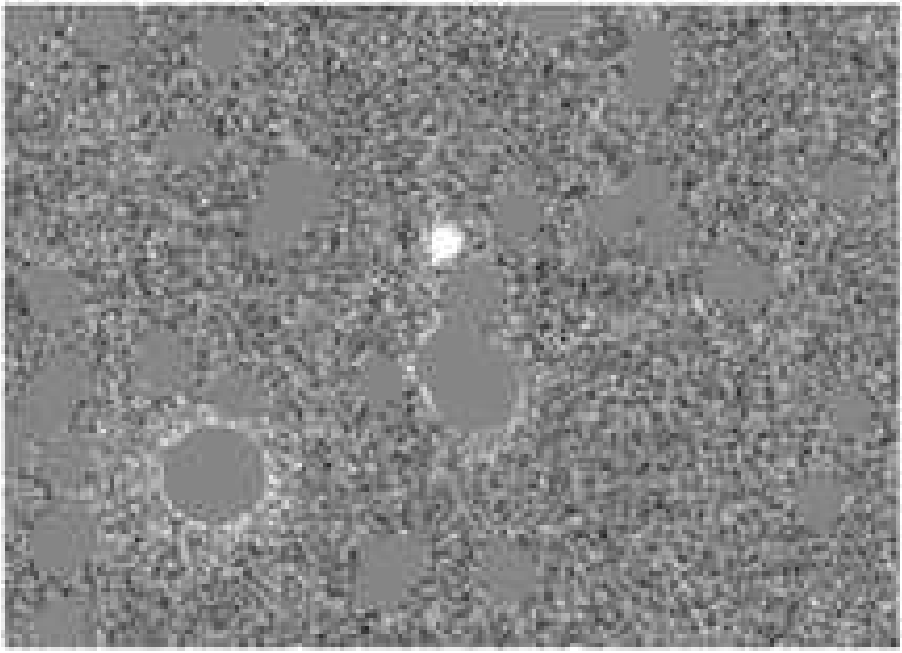}
\caption{A so-called ``Drop'' image produced by {\tt ConvPhot}
after the fitting procedure.  The name illustrates that each object
detected in the high resolution image, and hence fitted by {\tt
ConvPhot}, is multiplied by zero: this allows to identify objects in
the {\it measure} image (``drop''--out objects) that were not detected
in the {\it detection} one.}
\label{dropimage}
\end{figure}

{\tt ConvPhot} produces in output several quantities, including parameters for
each object as well as several type of images.

The most important output quantities are the best--fit solutions of
the minimisation procedure, i.e. the $F_i$ parameters and their relative
errors. Since the $P_i$
model for each object is normalised to unit flux, the resulting total
magnitude in the {\it measure} image is simply $-2.5 \log(F_i) + ZP_m$, 
where $ZP_m$ is the zero-point of the {\it measure} image itself. Based on 
our tests, we have concluded that this total magnitude is a reliable measure
of the actual total flux of the objects, somewhat less prone to systematic
effects than the Kron magnitudes computed by SExtractor. However,
these total magnitudes can be hardly compared with the SExtractor
magnitudes of the {\it detection} image, so that reliable colours cannot be 
directly obtained. A robust colour estimation, indeed, should be carried
out on a {\em same} area of the object and possibly be extended to a
large region of the source, in order not to be biased by red nuclei
(typical of bulges or elliptical galaxies) or by strong colour gradients,
especially in the rest frame UV wavelengths.
At this purpose, we use the total flux $D_i$ measured by 
{\tt ConvPhot} itself in the {\it detection} image as a good estimate for the
flux of the detected object, and used it to normalise the 
object profile $P_i$.
The resulting flux ratio is therefore 
$Flux(measure)/Flux(Detection) =
F_i / D_i \times 10^{-0.4(ZP_m - ZP_d)}$, 
where $ZP_d$ and $ZP_m$ are the zero-points of the {\it detection} and
{\it measure} image, respectively. 
This flux ratio, or the equivalent colour, 
$m_{measure}-m_{detection} = -2.5 \log(F_i / D_i) +ZP_m - ZP_d$
is the fundamental output of {\tt ConvPhot}. These useful quantities are
saved in the {\tt ConvPhot} output files, as well as the background
estimations,
the area used during the fitting procedure, the goodness of the fit and the
residuals left in the {\it measure} image.

In addition, {\tt ConvPhot} produces useful output images, such as the
model image
$M(m,n)$ with each object scaled to the fitted flux. In particular, we
produce a ``Residual'' image, containing the fit residuals, i.e.  the
quantity ${I(m,n)-B_i-M(m,n)}$. This image is useful to judge the
quality of the fit and the background estimation. Since large
residuals could typically occur in the center of bright objects, this image
is not very useful to perform an automated search of objects that
are detected only in the {\it measure} image but are not present in the
detection image (called ``dropout''). At this purpose,
{\tt ConvPhot} also produces a ``Drop'' image,
where all the pixels within the (dilated)
segmentation are set equal to zero. The example in Fig.~\ref{dropimage} is
taken from the K band image of the GOODS South field \citep{grazian}, where
a galaxy remains in the ``Drop'' image after the fitting of objects
selected in the $z$ band of ACS.


\section{Blended sources in the detection image}

The {\tt ConvPhot} algorithm is ideally suited for photometry of blended
objects on the {\it measure} image, but requires that the same objects in the
{\it detection} image should be well defined and their profiles are not
distorted
by noise. In the practical case, even in the GOODS-ACS images, there are
galaxies blended in the ACS images or faint objects, brighter than the
detection limit but still affected by noise in their shape/profile.
To overcome the risk of bad modelling for these sources, we introduce
in {\tt ConvPhot} an option to
carry out the fitting procedure only on the central part of the profile:
the fit is restricted only to the pixels where the detection image is
above a given relative threshold $t_f$.
This ensures that the fit is carried out only where the Signal to Noise
of the model ({\it detection} image) is high and avoids the contamination
from nearby sources,
which cannot be perfectly modelled in the detection image.

\begin{figure}[t]
\includegraphics[width=6.0cm]{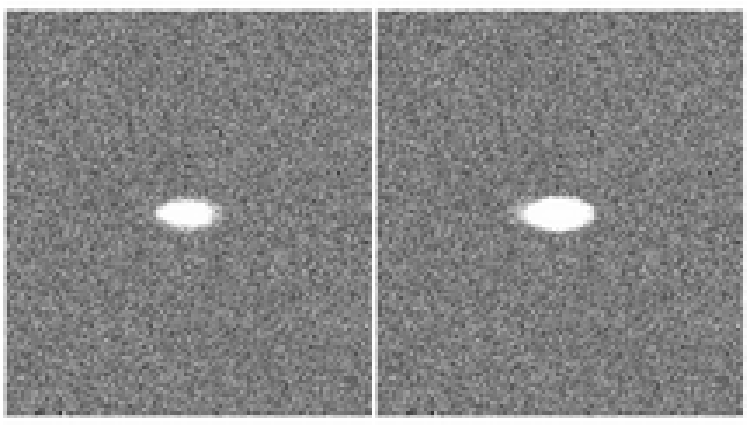}
\includegraphics[width=6.0cm]{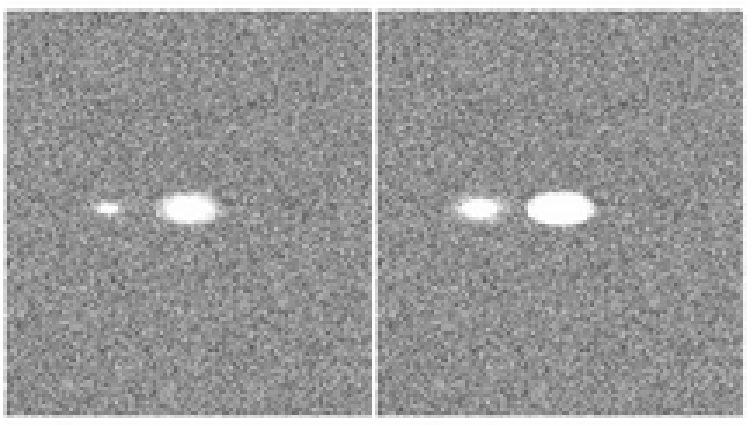}
\includegraphics[width=12cm]{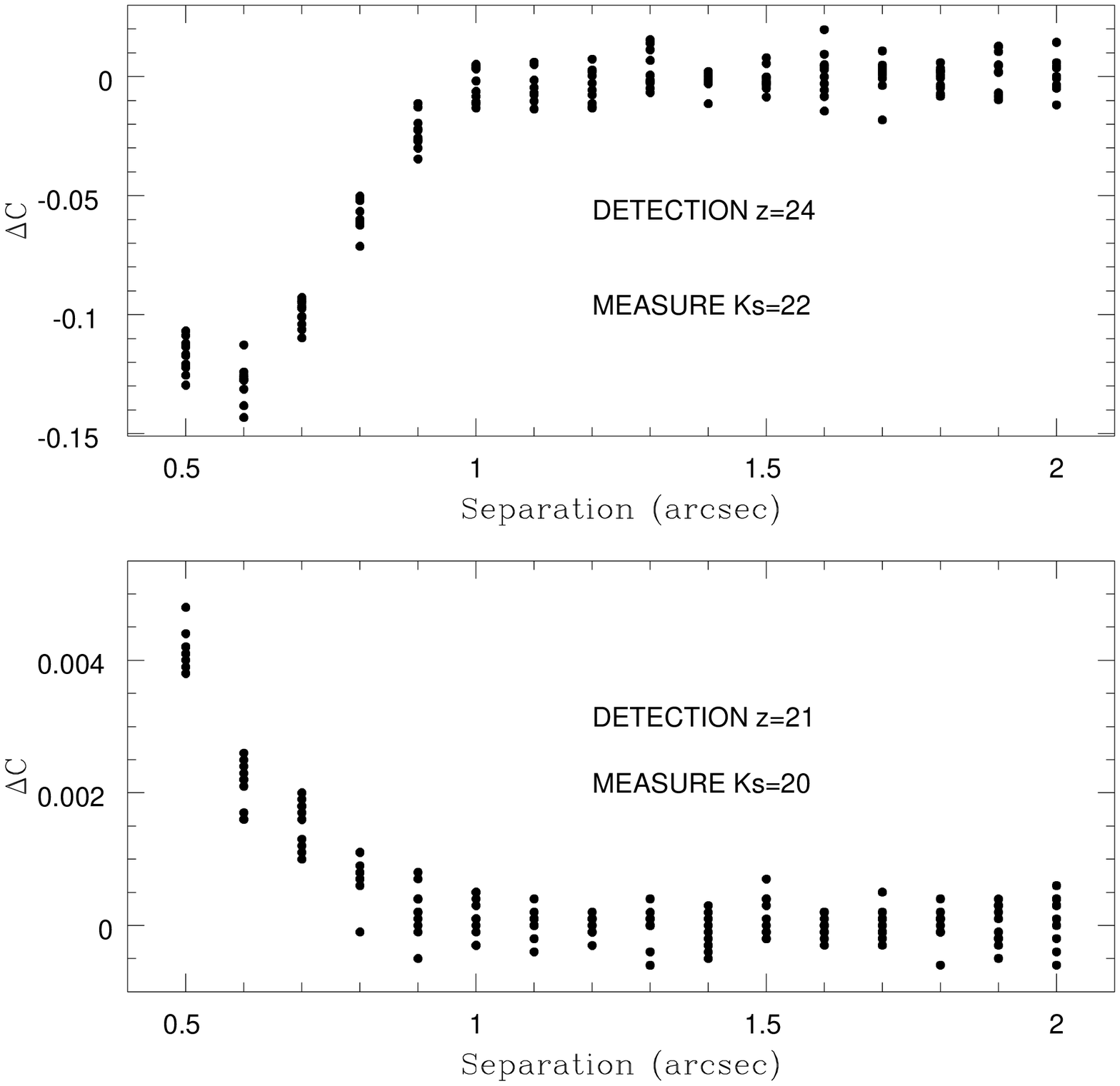}
\caption{
The effect of blending on the {\it detection} image in a standard
case. Two elliptical galaxies with $Rh=0.3$ arcsec are simulated as a
function of the angular separations in order to study the combined
effect of blending and colours. One object is a relatively bright,
blue galaxy ($z=21$, $z-Ks=1$), the second is a fainter, slightly
redder galaxy ($z=24$, $z-Ks=2$). The detection image has the
pixelscale and noise properties of the ACS GOODS images (seeing=0.12,
pixel scale=0.03 arcsec), while the {\it measure} image reproduces the
typical properties of a Ks band image of the GOODS data set, with a
FWHM seeing of 0.5 arcsec. For each
separation we computed 10 different realization of the galaxy
pairs. In the upper panel, bi-dimensional images of the galaxies with
a separation of 0.5 arcseconds (two leftmost images) and 2.0 arcseconds
(two rightmost images) are shown. In both cases, the first (left)
thumbnail is the $z$ band, while the second is the Ks band.  In the
lower panels we plot the difference between the input and derived
colours ($\Delta C$) for faint and bright sources as a function of the
angular separation. Even when the sources are deeply blended, the
error on the estimated colour for the fainter galaxy remains acceptable
($\simeq 0.1$ mags).  }
\label{blend_normal1}
\end{figure}

\begin{figure}
\includegraphics[angle=-90,width=12cm]{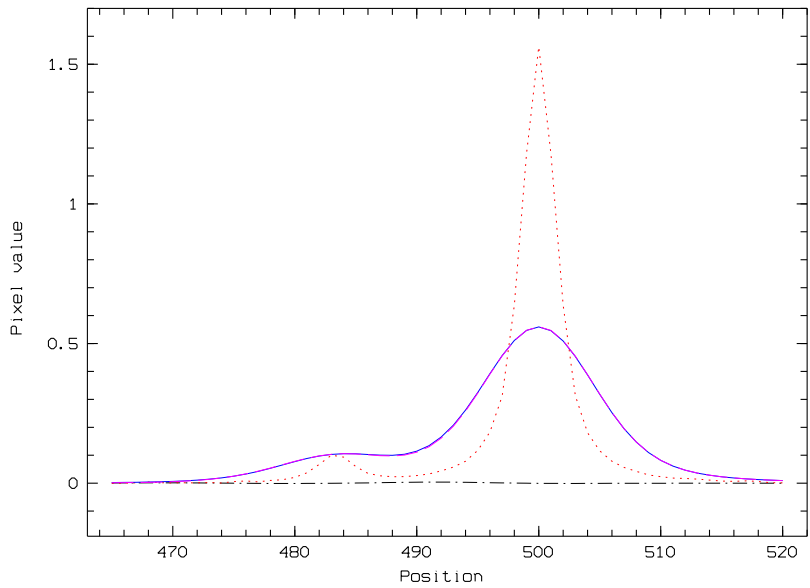}
\caption{
A blending case is simulated (upper left panel of
Fig.\ref{blend_normal1}), with distance of 0.5 arcsec between the two
peaks, for the objects described in Fig.~\ref{blend_normal1}.  The
position is given in pixel units and the pixel scale is 0.03 arcsec.
Solid lines shows the $Ks$ profile, dotted line the $z$ band profile,
long-dashed line the fit provided by {\tt ConvPhot} and the dot-dashed
line represents the residual. The fitting procedure succeeds in
recovering the true flux, as the flat residual demonstrates.
}
\label{blend_normal2}
\end{figure}

The effects of source blending on the {\it detection} image is
investigated by means of simulations. Fig.\ref{blend_normal1} and
Fig.\ref{blend_normal2} reproduce two elliptical galaxies (with half
light radius of 0.3 arcsec and ellipticity 0.5) with different angular
separations (between 0.5 and 2.0 arcsec) and show the difference
between the input colour and the output of {\tt ConvPhot},
$\Delta C=(z-Ks)_{out}-(z-Ks)_{inp}$.  The characteristics of the {\it
detection} image reproduces the $z$ band of ACS for the GOODS South
field (FWHM=0.12 arcsec, pixel scale=0.03 arcsec), while the {\it
measure} image fits the ISAAC $Ks$ band (FWHM=0.5 arcsec and pixel
scale of 0.03 arcsec to match the ACS frame). The RMS is chosen equal
to the typical value of GOODS $z$ band images (the $rms$ is 0.0014
ADU$/$Pixel), both for the {\it detection} and the {\it measure}
images and is added to the synthetic frames with the IRAF task {\tt
mknoise}.  The two galaxies are relatively bright in the $z$ band
(magnitude 21 and 24 in the AB photometric system), and have $z-Ks$
colours typical of local ellipticals ($z-Ks=1$ and 2).  We simulate 10
pairs of galaxies for each angular separation in order to have enough
statistics in the output quantities.  The colour of the brighter
source is recovered within few millimag even if the two sources
started to be blended in the detection image. The magnitude of the
fainter galaxy, however, is affected by light coming from the bright
neighbour, which produce a difference $\Delta C=-0.1$ in colour at small
angular separations ($\le 1^{''}$).

\begin{figure}
\includegraphics[width=6.0cm]{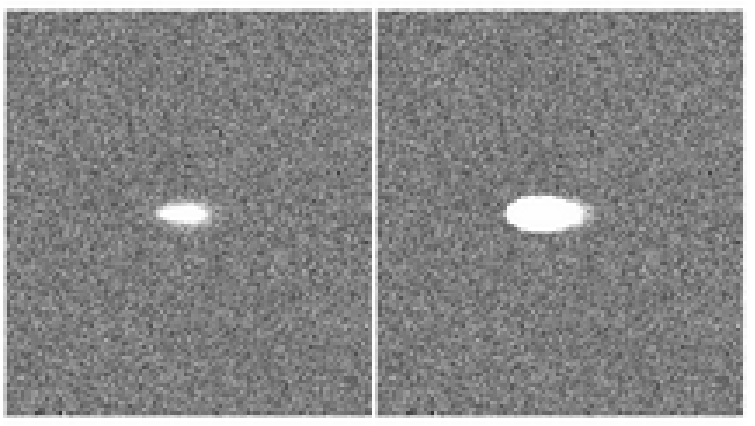}
\includegraphics[width=6.0cm]{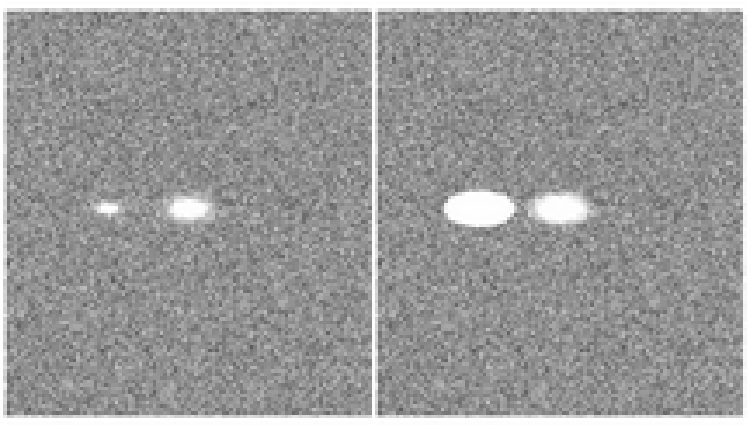}
\includegraphics[width=12cm]{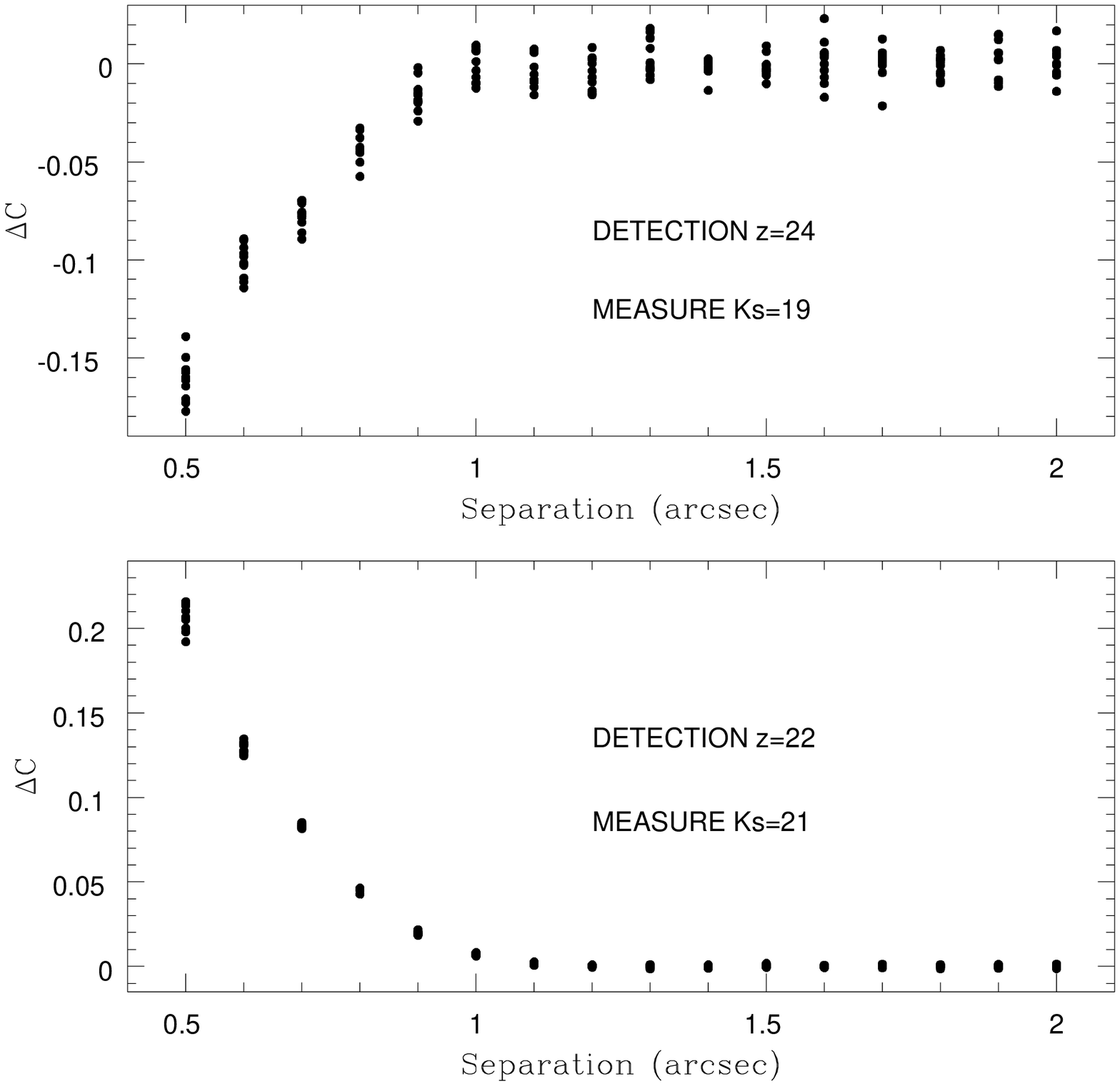}
\caption{
Simulations on the effect of blending on the {\it detection} image in
a case of extreme colours. Similar to Fig.\ref{blend_normal1}, two
elliptical galaxies with $Rh=0.3$ arcsec are simulated as a function
of the angular separations in order to study the combined effect of
blending and colours. At variance with Fig.\ref{blend_normal1}, colours
are so extreme here that the object that is fainter in the {\it
detection} image is indeed brighter in the {\it measure}
one. Specifically, one is a relatively faint, blue galaxy ($z=22$,
$z-Ks=1$), the second is a fainter but much redder galaxy ($z=24$,
$z-Ks=5$), in order to simulate an Extremely Red Object. {\it The
detection} image has the pixelscale and noise properties of the ACS
GOODS images (seeing=0.12, pixel scale=0.03 arcsec), while the {\it
measure} image reproduces the typical properties of a Ks band image of
the GOODS data set, with a FWHM seeing of 0.5 arcsec. For each
separation we computed 10 different realization of the galaxy
pairs. In the upper panel, bi-dimensional images of the galaxies with
a separation of 0.5 arcseconds (two leftmost images) and 2.0 arcseconds
(two rightmost images) are shown. In both cases, the first (left)
thumbnail is the $z$ band, while the second is the Ks band.  In the
lower panels we plot the difference between the input and derived
colours ($\Delta C$) for faint and bright sources as a function of
the angular separation. It is shown that the fit remains accurate as
long as the two objects are separated in the {\it detection} image
(i.e. for distances larger than 1.0 arcsec), although they are already
blended in the {\it measure} image. As the sources become deeply
blended in the{\it detection} image, with distance less than 1.0
arcsec, a non-negligible contamination starts to appear. Not
surprisingly, such contamination is larger for the bluer and fainter
(in Ks) galaxy.  }
\label{blend_eros1}
\end{figure}

\begin{figure}
\includegraphics[angle=-90,width=12cm]{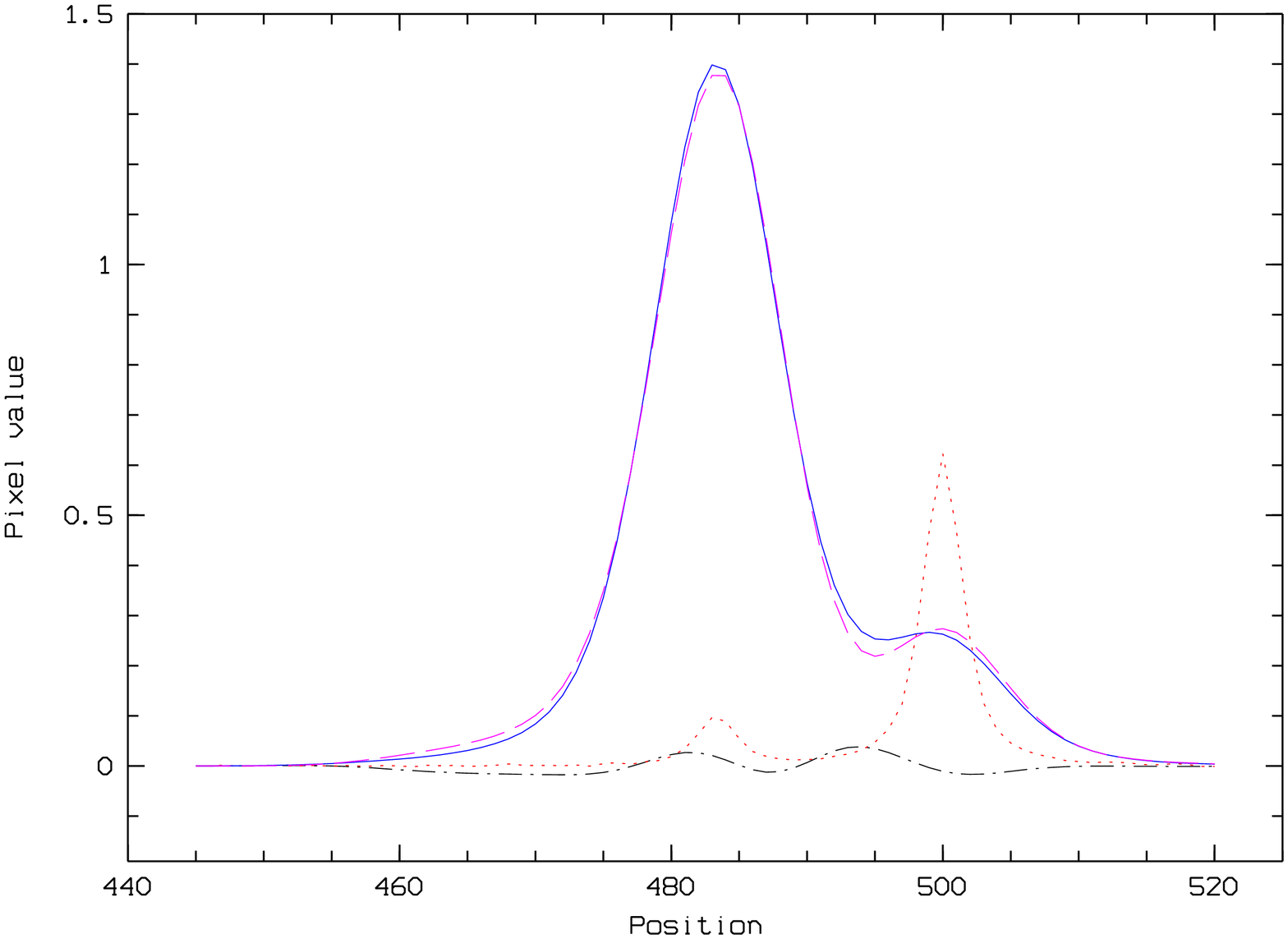}
\caption{
A deep blending case is simulated, with distance of 0.5 arcsec between
the two peaks, for the objects described in Fig.~\ref{blend_eros1}.
One object has an extreme colour ($z-Ks=5$).  The position is given in
pixel units and the pixel scale is 0.03 arcsec.  Solid lines shows the
$Ks$ profile, dotted line the $z$ band profile, the long-dashed line
the fit provided by {\tt ConvPhot} and the dot-dashed line the
residual.
}
\label{blend_eros2}
\end{figure}

Fig.\ref{blend_eros1} and Fig.\ref{blend_eros2}, indeed, describe the
effects of blending in an extreme case, where the fainter galaxy in
the detection image has a $z-Ks$ colour equal to 5, typical of an
Extremely Red Object (ERO), which reproduces the typical colours of
ellipticals at intermediate redshifts \citep{daddi}. In this critical
situation, the difference between the simulated and {\tt ConvPhot}
colour is 0.15-0.2 at a separation of 0.5 arcsec, and the fitting
procedure left visible residuals.  It is important to remark that the
fit remains accurate as long as the two objects are separated in the
{\it detection} image (i.e. for distances larger than 1.0 arcsec),
although they are already blended in the {\it measure} image. As the
sources become deeply blended in the{\it detection} image, with
distance less than 1.0 arcsec, a non-negligible contamination starts
to appear. Not surprisingly, such contamination is larger for the
bluer and fainter (in Ks) galaxy.  Fortunately, this situation is not
common even in large and deep areas like the GOODS survey, and even
for faint objects in the detection image the magnitude in the measure
image is recovered with acceptable accuracy. We have to stress that in
this case the parameters of the {\tt dilate} and {\tt ConvPhot}
softwares (dilation factor, minimum area, threshold, background
annulus) are not optimised for this special case but are the standard
values used for all the simulations in this paper.

\section{Systematics in the PSF-matching}
We briefly describe here the systematics that may affect such
PSF--matching procedure and the approaches that we have developed
within {\tt ConvPhot} to minimise them.

{\it Alignment errors}\\ The fitting procedure is extremely sensitive
to alignment errors.  In the simple case that the object has a 2--D
Gaussian shape of standard deviation $\sigma$ ($=FWHM/2.3548$), it is
easy to show that the resulting flux is {\it systematically
underestimated } by a factor $f = exp(-\frac{3}{4} \frac{\Delta
r^2}{\sigma^2})$, where $\Delta r$ is the offset in pixel of the
center position. When ground--based images are combined to HST images
with excellent sampling, this effect is non-negligible. In the case of
the GOODS ACS data, for instance, the ACS pixel-size is 0.03'', that
is often smaller than the residuals of the alignment of IR
ground--based images. For an alignment error of 1 pixel (in the {\it
detection} image), a figure that can be quite typical or even optimal
when combining ground--based and HST images, the resulting flux
underestimate may be of about 3\%.

As a first way out, we have included in {\tt ConvPhot} an option to
re-center any object before minimisation. In this case, the center of
each object are internally computed in the model as well as in the
measure image, and the measure is re-centered to the model image
before minimisation.  Since the center determination may be noisy for
faint objects, the user can set a limit on the S/N of the objects (in
each image) to execute this operation only on bright sources. We
remark that such option must be adopted with great care, after tests
and simulations on the specific data. In the case of the GOODS--MUSIC
catalog, for instance, we have adopted the recentering option for sources
with $S/N\ge 15$ both in the {\it model} and in the {\it measure}
image for the WFI U bands, while we decide not to use
this option for U-VIMOS, ISAAC and IRAC data.

{\it Variable FWHM or object profile}\\ Another source of uncertainty
may result from a variation of the object profile from the {\it detection} 
to the {\it measure} image. This can be due to either a physical change of 
the object profile (as due, for instance, to a more prominent bulge in the 
IR) or to an incorrect estimate of the PSF transformation kernel. In this
case, assuming that the profile of an isolated objects in the model
image and in the measure image are both Gaussian with same center and
different $\sigma$, the resulting flux is incorrectly estimated by a
factor $f=2
\sigma_{model}^2/(\sigma_{model}^2 + \sigma_{measure}^2)$. In this
case, the resulting flux can be therefore either under- or
over-estimated, depending on the sign of the error in the PSF
estimate. An error of 10\% in the object PSF will result in a 5\%
error in the output flux.

\begin{figure}
\includegraphics[angle=-90,width=12cm]{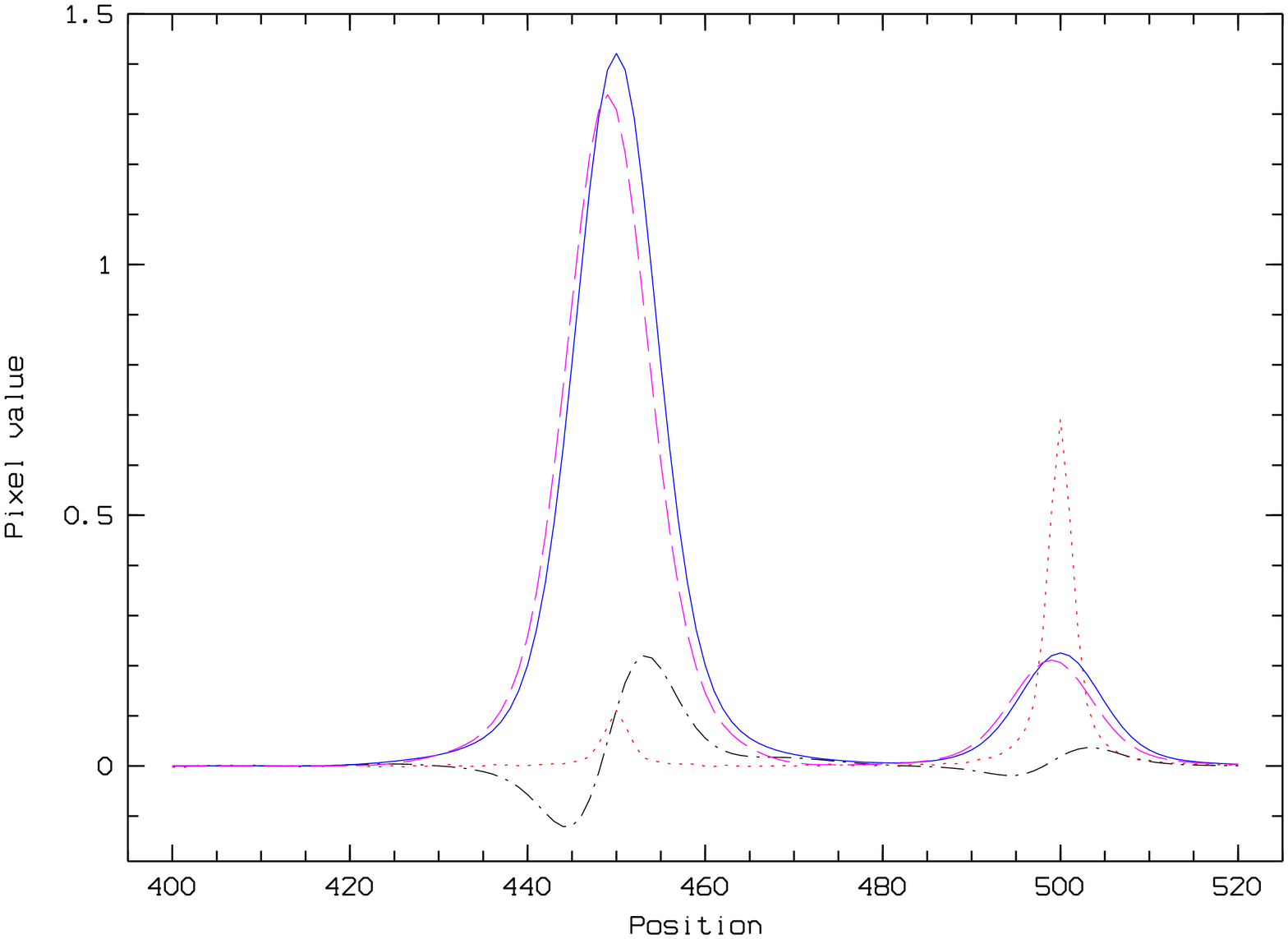}
\includegraphics[width=12cm]{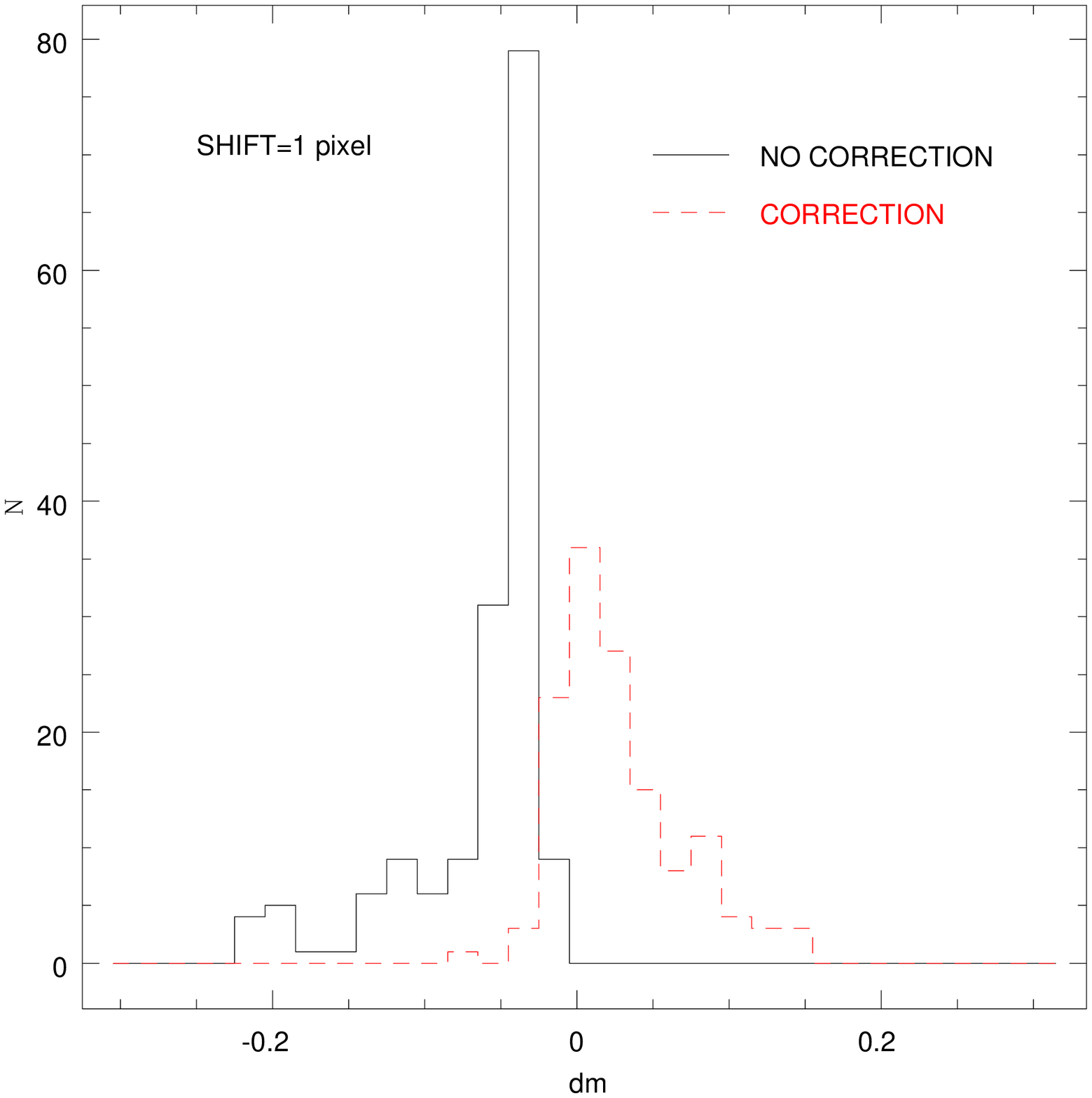}
\caption{The effect of not accurate centering of the detection and measure
images. The two objects are the same as in Fig. \ref{blend_eros1} but with
a distance of 2 arcsec (upper panel). The $z$ and $Ks$ images are shifted of
1 pixel. The position is given in pixel units and the pixel scale is 0.03
arcsec. The lower panel shows the difference between input and measured 
colours without correction (solid histogram) and taking into account the 
correction provided by {\tt ConvPhot} (dashed line).
}
\label{centering}
\end{figure}

The small systematic effects that have been described above, or others
resulting from different sources, can be efficiently corrected by
taking into account the flux in the residual image. At this purpose, we
have included in the code an option to compute the total residual flux
contained in the segmentation area of the original frame.
This correction is computed in a small region in order not to be
affected by nearby objects, which would drastically contaminate the
corrections computed in the residual image, especially for faint
objects severely blended with bright sources in the {\it measure} image. In
Fig.~\ref{centering} we show the efficiency of using the residuals to correct
possible biases due to centering problems.

{\it Objects with low Signal to Noise Ratio}\\
A well known drawback of our procedure occurs when the objects have a
low S/N in the detection image, since this would result in a noisy
determination of the profile, and hence of the final colour, even for
objects that have a high S/N in the {\it measure} image. During the
development of {\tt ConvPhot}, we indeed considered the possibility of
using some parametric fit as input shape of the objects. After some
test (which revealed also that the task is computationally heavy), we
realized that it was not a viable solution, for two reasons:
\begin{itemize}
\item 1)
a large fraction of faint galaxies have intrinsically irregular
morphologies, such that errors in their fitting produces at least
the same level of uncertainty or worse than using their observed profile,
albeit noisy;
\item 2)
for faint objects, the PSF-match smoothing
significantly reduces the noise, such that the leading error term is
anyway the global uncertainty on the total flux, that is also
present at the same level in the parametric fitting of the
shape.
\end{itemize}
 For these reasons, we did not insert such technique in {\tt
 ConvPhot}.


\section{Validation tests}

We have performed several validation tests on the {\tt ConvPhot} code,
during the debugging phase and to estimate the efficiency in the
correction for systematics.

\subsection{Test on simulated data}

A first, obvious set of simulations has involved the use of synthetic
images, obtained with the IRAF ARTDATA package, where we have
simulated either single or pairs of galaxies of various luminosities,
morphologies and resolutions, by which we have verified that the code
is computationally correct. Examples of this simulations are those
reproduced in Fig.\ref{blend_normal1} and Fig.\ref{blend_eros1} and
will not be presented here.

We have done more accurate tests to validate {\tt ConvPhot} when
dealing with the complexity of real data.  First of all, we checked
the reliability of {\tt ConvPhot} simulating the typical case of an
HST deep image as {\it detection} and a ground based image as {\it
measure}, using three cutouts of the real ACS images of the
GOODS-South field. As {\it detection} image, we have used the original
$z$ band (F850LP, 0.12 arcsec FWHM), with its relative segmentation
map and SExtractor photometry derived by the GOODS-MUSIC database
\citep{grazian}. As {\it measure} images, we have smoothed the 
$B$ (F435W) and $I$ band (F775W), with a 0.5 arcseconds kernel. We
have run {\tt ConvPhot} only on sources that are fully inside the {\it
measure} images ($f_{flux}=1$), expanding the thumbnails of 128 pixels
in order to have enough pixels to compute the background with {\tt
ConvPhot}, and with threshold 0.0 and 0.5 (with the latter choice we
fit only the central part of the objects).  

\begin{figure}
\includegraphics[width=12cm]{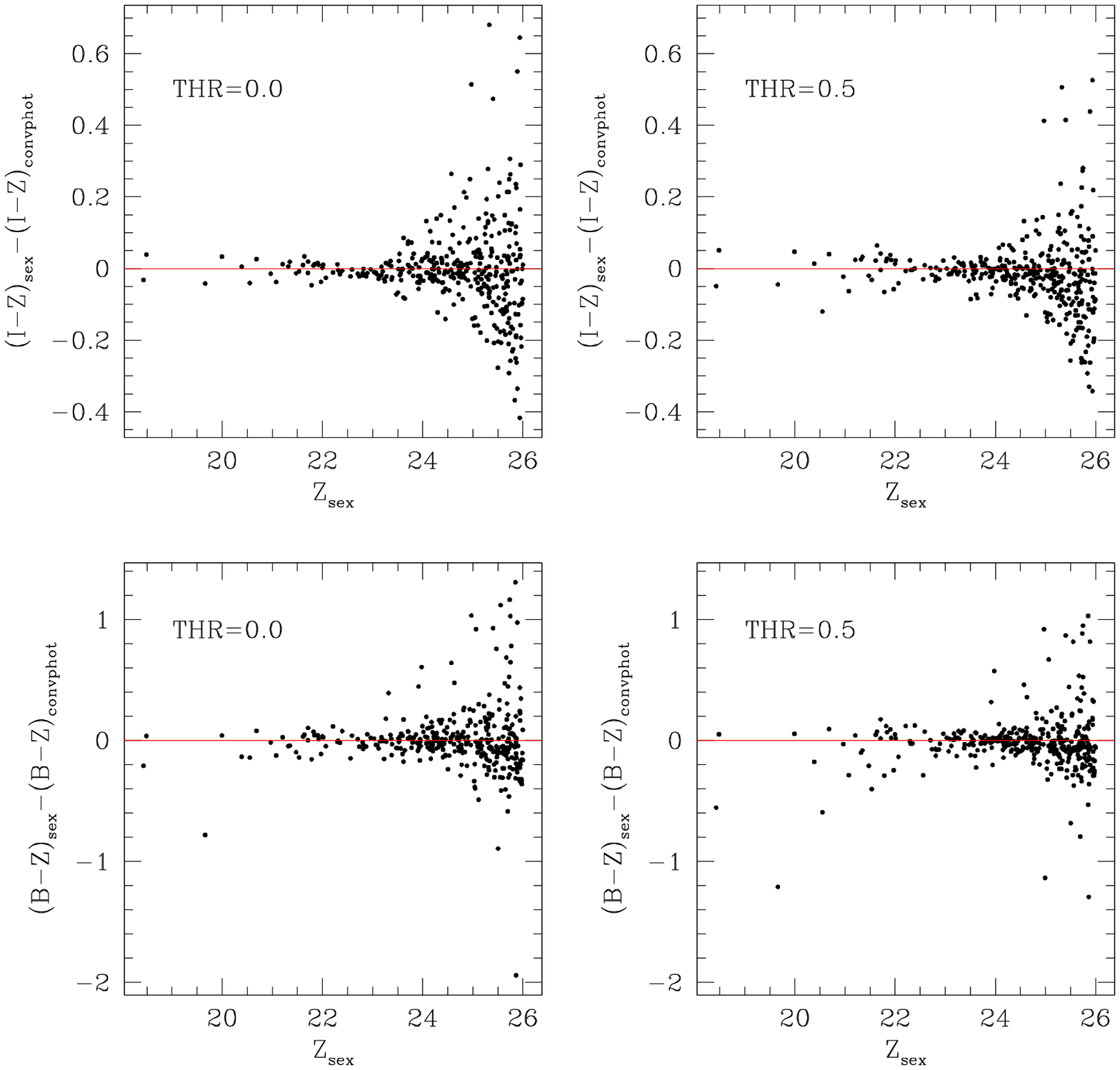}
\caption{The $I-Z$ and $B-Z$ colours as a function of the $z$ band magnitude
of the objects detected on the GOODS-South survey. Two different values of the
threshold are used (0.0 and 0.5) and they affect in dissimilar way the
bright and faint objects. The small scatter in the input and ConvPhot colours
shows that the {\tt ConvPhot} colour determination is not
biased and is compatible with aperture photometry.
}
\label{bz_iz}
\end{figure}

As a figure of merit, we compare in Fig. \ref{bz_iz} the $I-Z$ and
$B-Z$ colours obtained by {\tt ConvPhot} with the aperture colour
obtained by SExtractor on the original, high resolution images, that
are those used in \citep{grazian}.

\begin{figure}
\includegraphics[width=12cm]{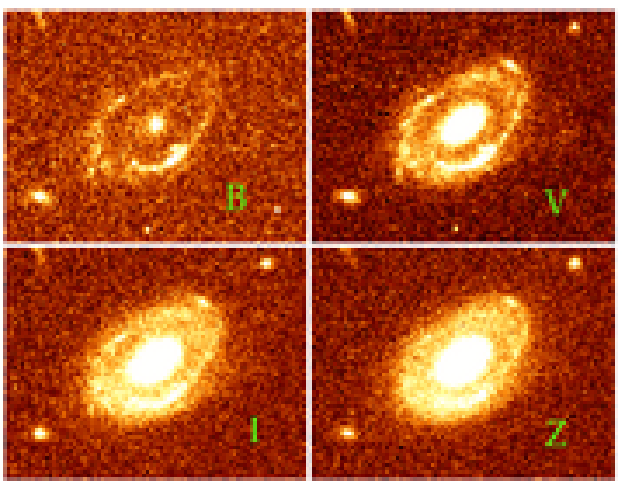}
\caption{A nearby galaxy in different optical bands (B,V,I,Z) taken from
the GOODS-MUSIC database. The different morphologies indicate the predominance
of star-forming regions in the B band in a ring around the galaxy disk and
the compactness of the old stellar population in the Z band in the bulge.
}
\label{obj}
\end{figure}

At first glance, this plot shows that our software is not biased in
the colour determination for bright as well as for faint sources.
Looking into the details, the residuals of the colours show a
different behaviour for bright ($Z_{sex}\le 23$) and faint galaxies.
The residuals of the bright sub-sample have larger dispersion both in
the B band and in the 0.5 threshold cases: an investigation by eye of
these bright sources with large scatter in the colour residuals
indicates that the main reason is the morphological change of source
profile for nearby bright galaxies, as for example that in
Fig. \ref{obj}.  The profiles of nearby and bright galaxies differ
largely when observed in the near UV and in the near IR, since in the
UV the star forming regions are clearly visible, and their spatial
distribution is significantly different than the regions occupied by
old/evolved stars, which emit mainly in the near IR. In this case, if
the galaxy profile in the $Z$ band is used to derive the magnitude of
the source, it will result in a slightly poorer fit, especially when
the wavelength different is large (B vs Z bands) or when the fit is
restricted to the small central bulge (threshold=0.5).

For faint sources, however, the difference between the $B-Z$ and $I-Z$
is not significant, while an higher threshold is useful to enhance the
quality of the fit and the accuracy of the magnitude estimation.

\subsection{Real data}

\begin{figure}
\includegraphics[width=12cm]{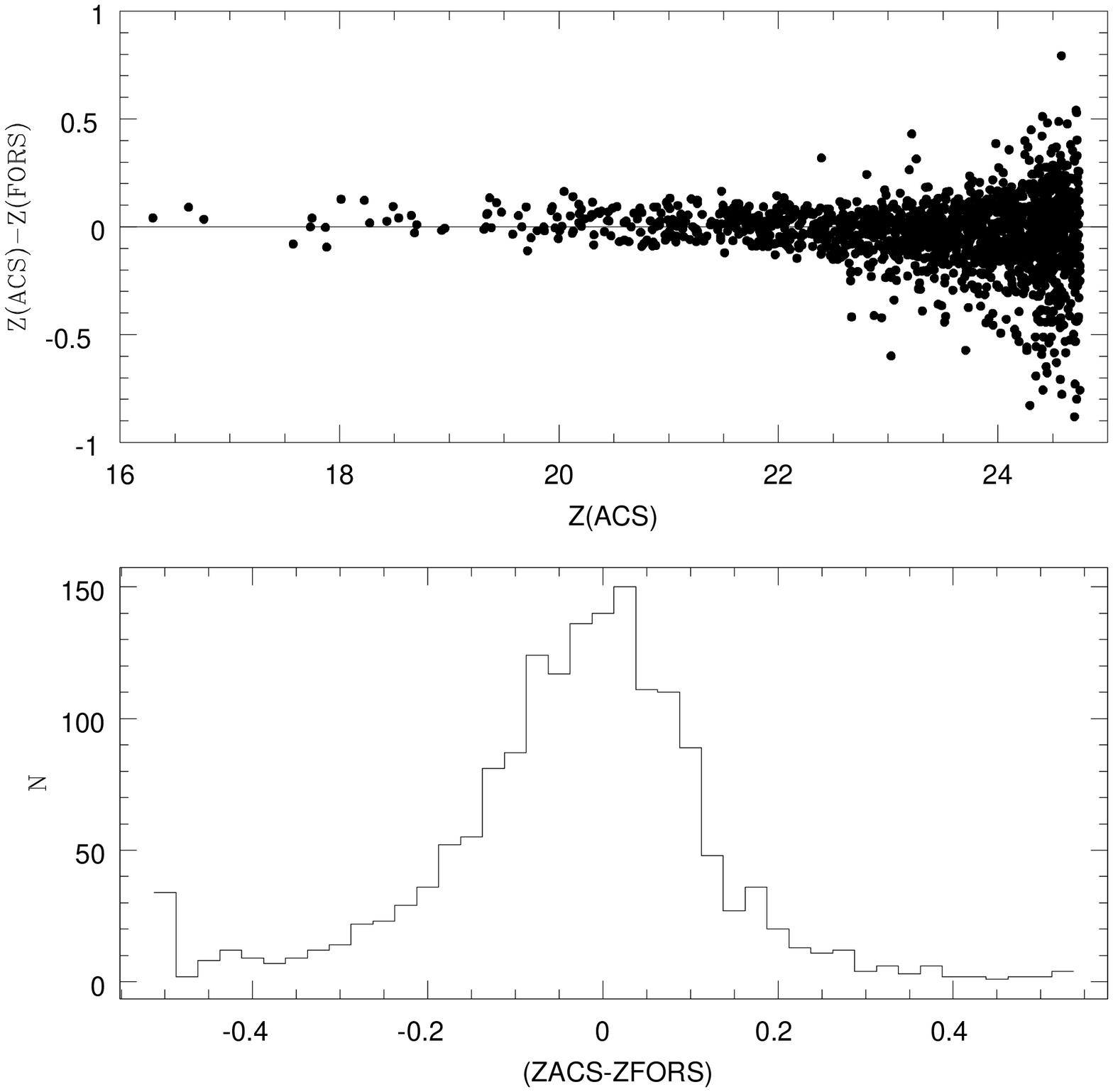}
\caption{The $z_{ACS}-z_{FORS}$ colour as a function of the magnitude of the
objects detected on the K20 survey (upper panel). The histogram of this
colour (lower panel) shows that the {\tt ConvPhot} colour determination is not
biased and is compatible with aperture photometry.
}
\label{FORS}
\end{figure}

Simulations, however, cannot reproduce the complexity of real objects
and data. To obtain a more stringent and independent test we have made
use of the $z$ band FORS image of the K20-CDFS (seeing=0.55
arcsec and pixel scale=0.2 arcsec). We have already analysed the K20
data set in a previous paper (Cimatti et al 2001), applying a standard
technique based on aperture photometry obtained with SExtractor. Here,
we have used {\tt ConvPhot} to obtain a new estimate of the
$z_{ACS}-z_{FORS}$ colour in the FORS image of the K20, and compared
them with the previously published catalog.  To this aim, we have
used the GOODS ACS z-band image as {\it detection}, with a typical
FWHM of 0.12 arcsec and pixel scale of 0.03 arcsec. The results of
this test are shown in Fig.\ref{FORS}, which are not corrected for the
small offset in the average colour ($<z_{ACS}-z_{FORS}>=0.05$) due to
the small difference in the filter response curve. Once allowing for
this offset, the plot shows that {\tt ConvPhot} is not biased in the
colour estimate in respect to standard aperture photometry.

\subsection{An extreme example: the case of MIPS data}

We have also made extensive tests in a case when {\tt ConvPhot} is
pushed to its very limits, i.e. when the resolutions, pixel scales and
wavelengths of the two images are very different. Such a situation is
very challenging both conceptually and technically, since the present
version of {\tt ConvPhot} works on {\it detection} and {\it measure}
images after aligning and rebinning them to the same pixelscale and
reference frame (a discussion on future possible improvements is given
in the following).  At this purpose, we have taken the extreme case of
applying {\tt ConvPhot} to a combination of HST--ACS and Spitzer--MIPS
images which have pixel scales of 0.03 and 1.2 arcsec/pixel,
respectively. Such data set is again taken from the public data of the
GOODS survey, where we take again the ACS $z$ band as {\it detection}
image and the MIPS image at $24\mu m$ as {\it measure}.

In principle, we should align and resample the original MIPS images,
with a typical PSF of 5.2 arcsec, at the same scale of the ACS
ones. This would result in a minimum size of the resulting thumbnail
of $2000\times2000$ pixels, that would make impracticable the use of
{\tt ConvPhot} even with fast workstations. To circumvent this
problem, we rebinned the ACS {\it detection} image by a factor
$8\times8$ (0.24 arcsec/pixel), resulting in a severely undersampled
image (the FWHM of the ACS GOODS image is 0.12 arcsec), reducing
significantly the image size. We also rebinned of the same amount
(with a specific code) the input Segmentation image. Finally, we
aligned and rebinned the MIPS data to such a rebinned ACS image. The
advantage of this procedure is that the requirements of memory and CPU
time are significantly reduced, although we loose some information
when introducing a undersampling of the original image.

\begin{figure}
\includegraphics[width=12cm]{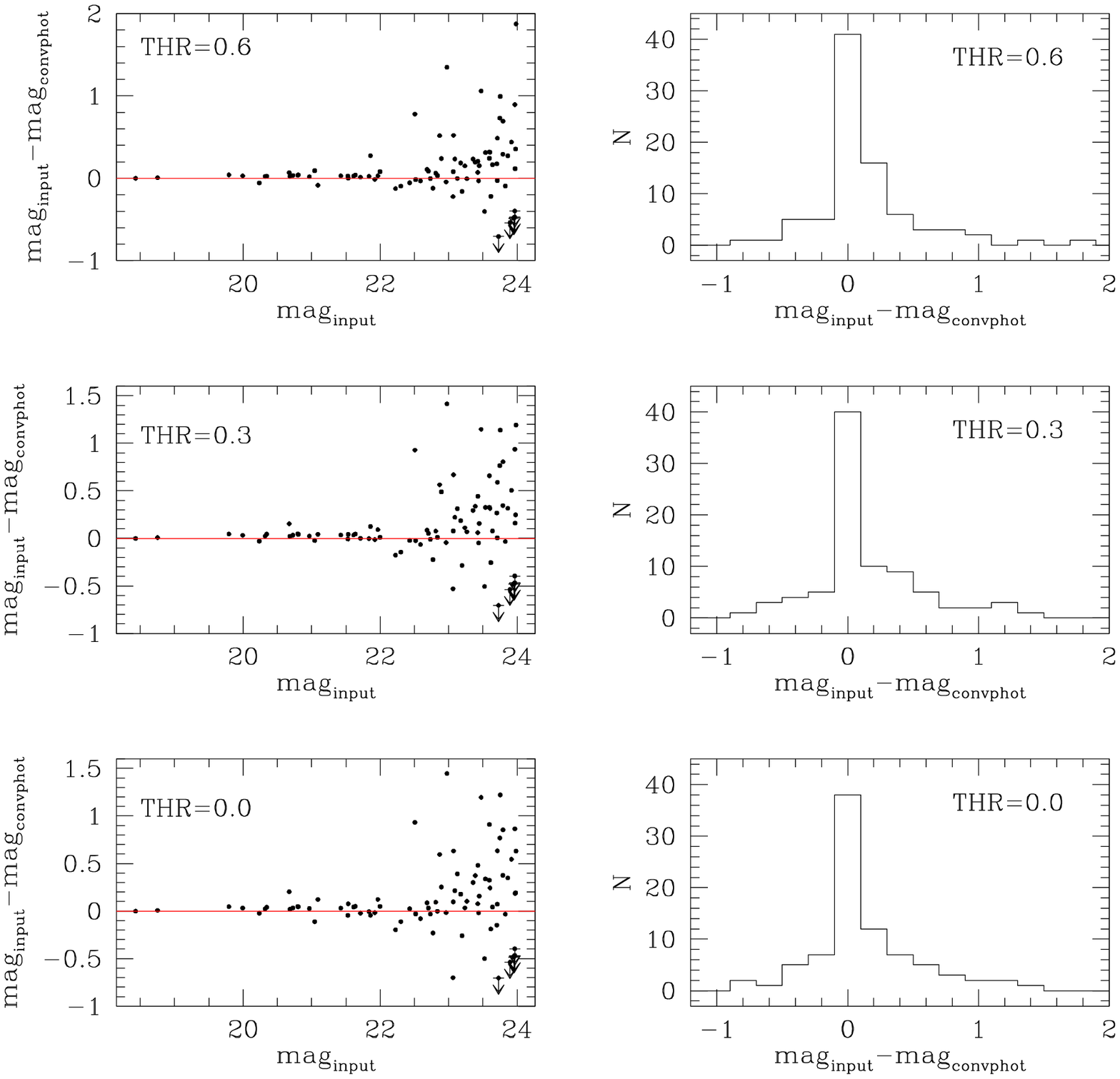}
\caption{Results of the simulations which reproduce the $z$ band of
ACS and the 24 micron band of MIPS
as {\it detection} and {\it measure} images, respectively. The y--axis
shows the difference between the input magnitudes and those derived by
{\tt ConvPhot} as a function of the magnitude on the simulated
objects. The simulated image is derived by smoothing the original $z$
band, so that the expected colour should be zero. The histogram of the
colours (right panels) shows the result for different value of the
threshold in {\tt ConvPhot}. Arrows indicate upper limit in the {\tt
ConvPhot} magnitude.  }
\label{zmipssim}
\end{figure}

Before applying it to the real data, we performed a simulation by
smoothing the rebinned ACS image to the same PSF of the MIPS data, and
used {\tt ConvPhot} to estimate the colour between the original and the
MIPS--smoothed one.  Fig. \ref{zmipssim} shows the comparison between
the input and the {\tt ConvPhot} magnitudes, indicating that this
software can be used even when dealing with such extreme cases.  This
plot shows also that the thresholding option for {\tt ConvPhot} can be
useful to reduce the scatter of the magnitude estimation.

We finally obtained a MIPS catalog of all sources in the GOODS--MUSIC
sample, after rebinning the original ACS z-band detection and
segmentation images by a factor of 8, computing the appropriate
kernel and run {\tt ConvPhot} over the entire GOODS field.  We have
used a threshold of 0.3 and an $f_{flux}$ factor of 0.2, on the
indications of the simulations described in the previous section.

To check the reliability of our magnitude estimation in the real MIPS
image, we compare the {\tt ConvPhot} magnitude estimate with those
derived by SExtractor. These are computed on circular apertures of 6
and 6.6 arcsec radius, and then corrected applying an appropriate correction
factor given in the MIPS Instrument Manual\footnote{\it
http://ssc.spitzer.caltech.edu/mips/}, in this case 0.58 and 0.54
magnitudes, respectively. The corrected aperture magnitudes are an
unbiased estimate
of the true source luminosity only if the object is
isolated. Obviously, given the large FWHM of the MIPS 24 micron Point
Response Function (or PSF), the effect of crowding on the aperture
magnitude is severe for faint sources, typically at $mag\ge 19$.
 
\begin{figure}
\includegraphics[width=6cm]{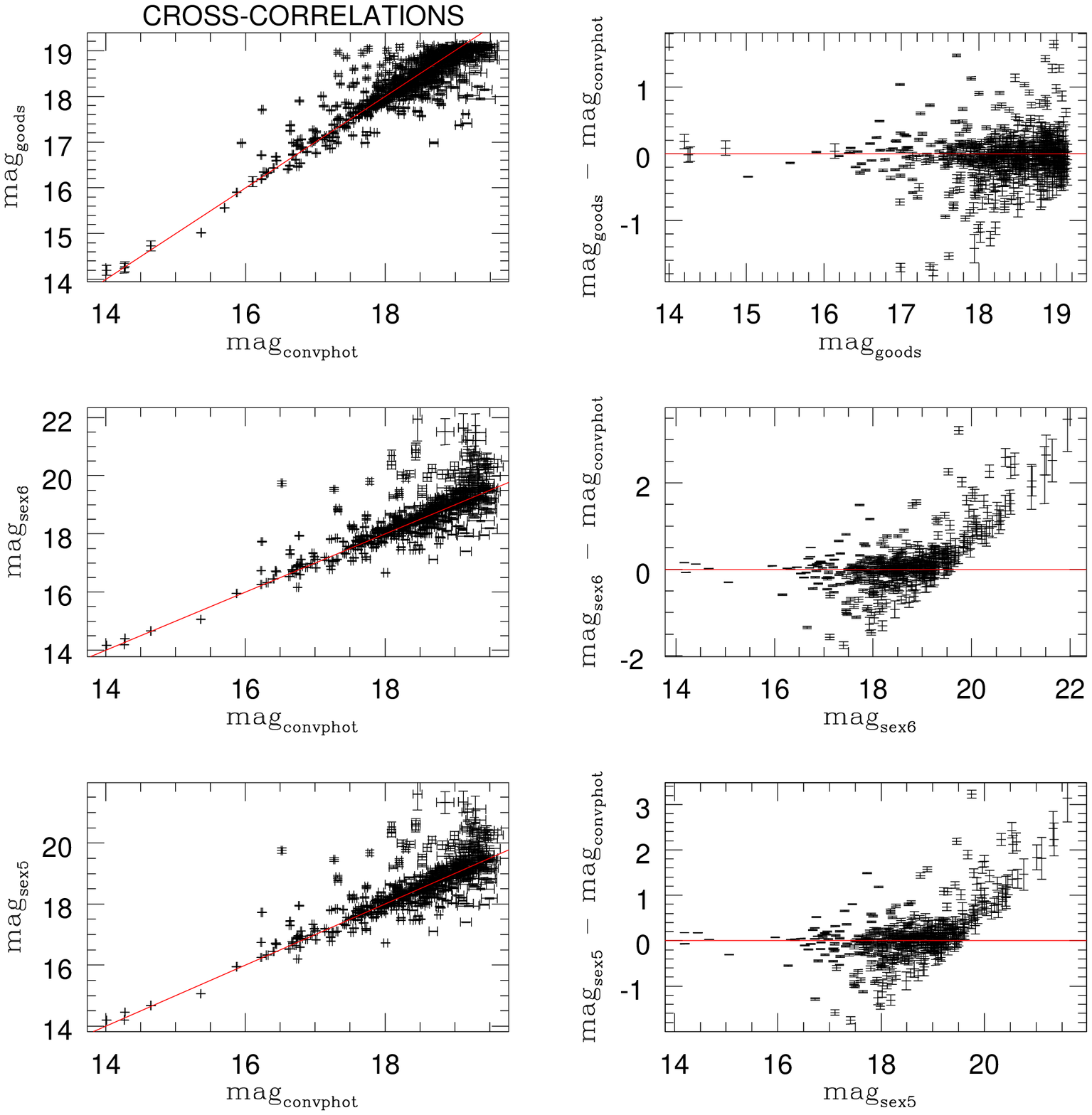}
\includegraphics[width=6cm]{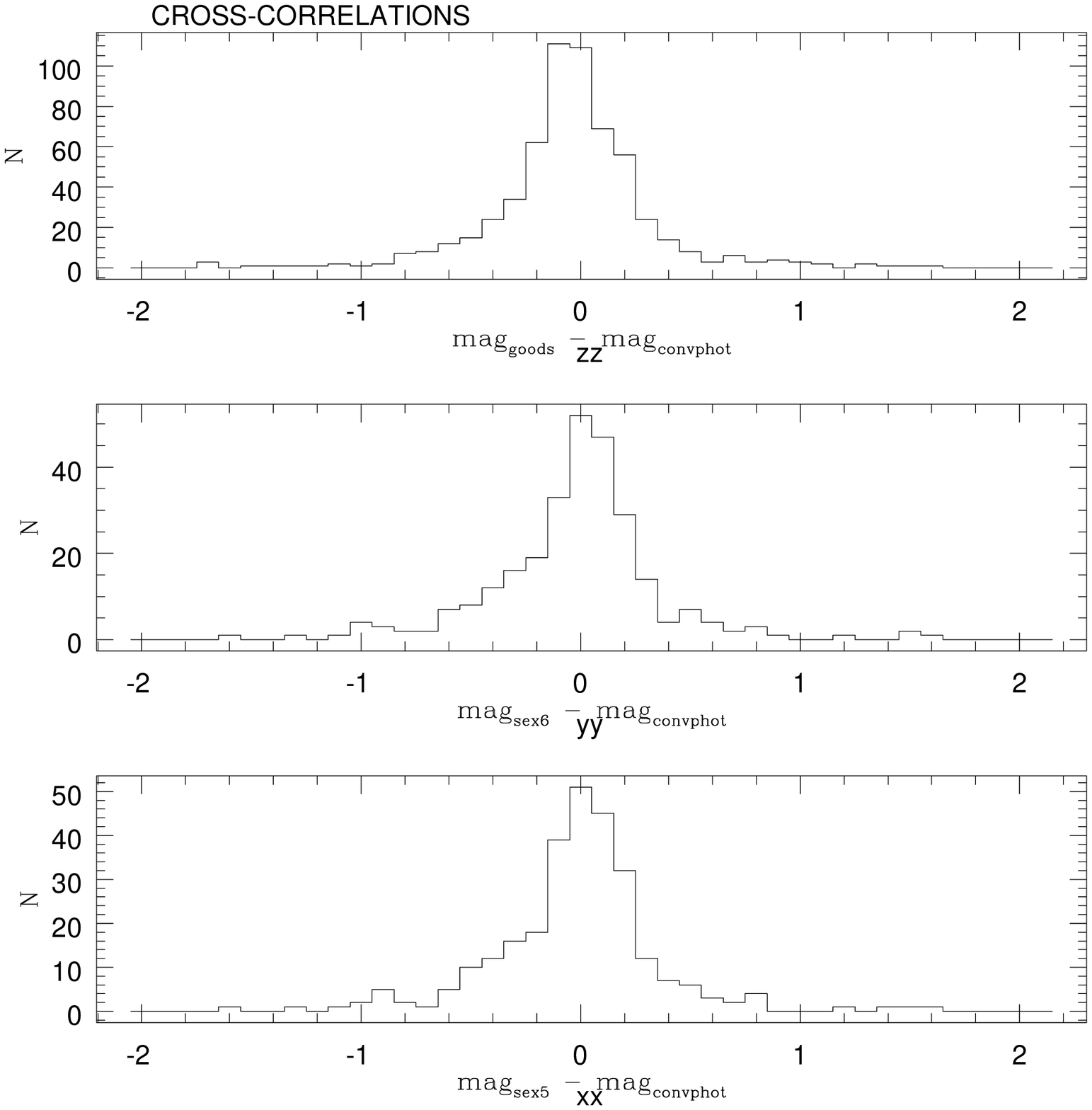}
\caption{The difference between the SExtractor aperture magnitude, corrected
for the flux fraction missed by the small circular aperture, and the
magnitude derived by {\tt ConvPhot} for the MIPS 24 micron sources in
the GOODS-South field. The histogram of
the magnitude residuals shows that the {\tt ConvPhot} colour determination
is not biased, while the comparison of the {\tt ConvPhot} magnitudes
with those derived by \cite{chary} shows a small (0.1 mag) offset.
}
\label{mipsreal}
\end{figure}

Fig. \ref{mipsreal} shows the comparison between the corrected
aperture magnitudes and those derived by {\tt ConvPhot} for the entire
GOODS-South field, indicating that even in the real case our algorithm
retrieve the correct fluxes. If a comparison is made with the
photometric catalog released by the GOODS Team \citep{chary}, who used
a similar technique starting from the IRAC 3.6 micron as {\it
detection} image, we find a good agreement for relatively bright
objects. This indicates that our background estimation is correct and
the {\tt ConvPhot} result is robust.  After the comparisons both with
simulations and real data, we are thus confident that the {\tt
ConvPhot} algorithm is stable even in this extreme case, both due to
the wavelength and PSF differences between the {\it detection} and
{\it measure} images.

A final remark is necessary about the use of undersampled image. We
have made a test using undersampled images (obtained by rebinning the
usual GOODS--$z$ band image by a factor 8), and simulated a ground
based image with seeing of 0.5 arcsec. In this case, we found evidence
that the {\tt ConvPhot} magnitudes are significantly biased, with a
typical underestimate of the fitted flux by 0.2 magnitudes. We
therefore conclude that - not surprisingly - the use of undersampled
images as {\it detection} images should be avoided when the difference
in resolution and pixelscale is less extreme than the MIPS case
described above..

This is just an example of how much the use of this software however
is delicate. We therefore strongly recommend the interested user to
make a set of realistic simulations to check the effect of the
multiple parameters on the correct magnitude estimation.

\section{Known ConvPhot issues: room for improvement}

The present version of {\tt ConvPhot} has several limitations, which
are described here and will be solved in future versions of this
software.  We summarise the most important in this brief list, and
defer to future version of the Code their implementation.
\begin{itemize}
\item
A major limitation results from the fact that all the thumbnails of
the processed sources are stored in the computer memory. This allows
to carry out the computation faster but limits the size of the images
on which {\tt ConvPhot} operates.
\item
The {\it measure} image and its RMS must be rebinned to the {\it detection}
image reference frame. In this case the alignment between sources can be
done accurately using dedicated routines (i.e. swarp, IRAF) and the advantage
of dealing with small pixel scales is to carry out a fine tuning of the
source centering in {\tt ConvPhot}. A draw back is that it increases the
thumbnail dimensions and consequently reduce the number of sources to be
stored in the computer memory. A possible way out would be to avoid at all
any rebin, and work directly in astrometric coordinates. 
\item
If two sources are blended in the {\it detection} image their profiles
in the model image will be different from the real case. A modelling of the
object profile for blended sources is definitely needed, but its implementation
is not easy, especially for irregular sources.
\item
The convolution with the input kernel is performed in pixel space, which is
computationally an heavy process. A convolution in the Fourier space would
allow a much faster computation.
\item
The kernel derivation is left to the user and it is fixed for all the
sources in an image. It is not foreseen for the moment the possibility of
compute the smoothing kernel internally or using a PSF which varies according
to the position of the objects.
\item
The recentering option is very simple, since it takes the maximum of a source
in the {\it detection} and {\it measure} image and compute the relative shifts.
This is not accurate for the case of crowded fields or large PSF images, in
which a mis-identification of two close sources turns out in a wrong
magnitude estimation.
\item
The option to correct the fitted flux with a statistic of the residual
image in the original segmentation is not satisfactory when dealing
with large kernel, and more simulations are necessary to find a proper
solution for these cases.
\end{itemize}


\section{Summary and Conclusion}

We have described in this paper a new, public software for accurate
multiband photometry of images of different resolution and depth, that we
have named {\tt ConvPhot}. The code is based on an algorithm that has been
originally proposed by FSLY99 and applied in some
previous analysis of
deep extragalactic imaging data \citep{lanzetta,papovich}.
The code performs a
``PSF-matched'' photometry on two images with different PSF, by first
extracting all objects from the first (detection) image, smoothing each of
them to the PSF of the second (measure) image, and obtaining the flux of each
object on the measure image by a global $\chi^2$ minimisation over the
whole image.

We provide here for the first time a public version of such a code, of
which we discuss the basic algorithm, the possible systematic effects
involved and the results of a set of simulations and validation tests
that we have performed on real as well as simulated images.  To
maximise the usability of the code, we explicitly use the outputs of
the popular SExtractor code as inputs to our procedure.

This code has been extensively used and tested to obtain the
GOODS--MUSIC sample, based on the public release of the GOODS--South
data, as we have described in Grazian et al. (2006). In such a
context,  we used {\tt ConvPhot} to obtain colours for more than 18000
sources originally detected in a wide, high resolution mosaic of
HST-ACS images. It has proven to be reliable enough to flawlessly
process around 70 frames at different wavelengths and image
quality and depth, ranging from VLT--VIMOS images in the U band (with
typical PSF of 1'') to VLT--ISAAC images in J, H and K (with typical
PSF of 0.5''), as well as Spitzer-IRAC images (with typical PSF of 2''). We
believe that the number of options in the code make it flexible enough
for a wider use.

{\tt ConvPhot} is written in C language under the GNU Public License and is
publicly released worldwide.  This software is described
in detail in a WEB site where the source code can be retrieved,
together with the user manual: {\tt
http://lbc.oa-roma.inaf.it/convphot}.

\subsection*{Acknowledgements}
We warmly thank the referee, M. Strauss, for his useful
suggestions and comments, that largely improved the quality and amount
of information in this paper.  It is a pleasure for us to warmly thank
Giulia Rodighiero for her patience during the software testing/usage
and for very useful comments to improve this paper, the {\tt ConvPhot}
software and its user manual.

\end{document}